\documentclass[fleqn,usenatbib,useAMS]{mnras}

\usepackage{graphicx}	
\usepackage{amsmath}	
\usepackage{multicol}        
\usepackage{bm}		
\usepackage{pdflscape}	
\usepackage{latexsym}
\usepackage[authoryear]{natbib}
\usepackage{bm,nicefrac}

\usepackage[T1]{fontenc}
\usepackage{ae,aecompl}

\usepackage{newtxtext,newtxmath}

\title[GMOS transmission spectroscopy of WASP-121b]{Gemini/GMOS Optical Transmission Spectroscopy of WASP-121b: signs of variability in an ultra-hot Jupiter?}

\author[J. Wilson et al.]{Jamie Wilson$^{1}$\thanks{Contact e-mail: \href{mailto:jwilson34@qub.ac.uk}{jwilson34@qub.ac.uk}}, {Neale P. Gibson$^{2}$}, {Joshua D. Lothringer$^{3}$}, {David K. Sing$^{3,4}$},\newauthor
{Thomas
 Mikal-Evans$^{5}$}, {Ernst J. W. de Mooij$^{1}$}, {Nikolay Nikolov$^{6}$} and {Chris A. Watson$^{1}$}
\smallskip
\\
$^{1}$Astrophysics Research Centre, School of Mathematics and Physics, Queens University Belfast, Belfast BT7 1NN, UK\\
$^{2}$School of Physics,  Trinity College Dublin, The University of Dublin, Dublin 2, Ireland\\
$^{3}$Department of Physics and Astronomy, Johns Hopkins University, Baltimore, MD, USA\\
$^{4}$Department of Earth $\&$ Planetary Sciences, Johns Hopkins University, Baltimore, MD, USA\\
$^{5}$Kavli Institute for Astrophysics
 and Space Research, Massachusetts Institute of Technology, Cambridge, MA 02139, USA\\
 $^{6}$Space Telescope Science Institute, 3700 San Martin Drive, Baltimore, MD 21218, USA}

\date{Last updated 2021 March 9; in original form 2021 March 9}

\pubyear{2021}

\begin{document}
\label{firstpage}
\pagerange{\pageref{firstpage}--\pageref{lastpage}}
\maketitle

\begin{abstract}
We present ground-based, spectroscopic observations of two transits of the ultra-hot Jupiter WASP-121b covering the wavelength range $\approx$\,500\,--\,950\,nm using Gemini/GMOS. We use a Gaussian process framework to model instrumental systematics in the light curves, and also demonstrate the use of the more generalised Student's-T process to verify our results.
We find that our measured transmission spectrum, whilst showing overall agreement, is slightly discrepant with results obtained using HST/STIS, particularly for wavelengths shortward of $\approx$\,650\,nm. In contrast to the STIS results, we find evidence for an increasing blueward slope and little evidence for absorption from either TiO or VO in our retrieval, in agreement with a number of recent studies performed at high-resolution. We suggest that this might point to some other absorbers, particularly some combination of recently detected atomic metals, in addition to scattering by hazes, being responsible for the excess optical absorption and observed vertical thermal inversion.
Our results are also broadly consistent with previous ground-based photometry and 3D GCM predictions, however, these assumed different chemistry to our retrievals.
In addition, we show that the GMOS observations are repeatable over short periods (days), similarly to the HST/STIS observations. Their difference over longer periods (months) could well be the result of temporal variability in the atmospheric properties (i.e. weather) as predicted by theoretical models of ultra-hot Jupiters; however, more mundane explanations such as instrumental systematics and stellar activity cannot be fully ruled out, and we encourage future observations to explore this possibility.

\end{abstract}

\begin{keywords}
methods: data analysis, stars: individual (WASP-121), planetary systems, techniques: Gaussian processes, techniques: spectroscopic
\end{keywords}



\begingroup
\let\clearpage\relax

\endgroup
\newpage

\section{Introduction}
Close-in giant exoplanets are amongst the most promising targets for detailed atmospheric studies of planets outside our own Solar System, owing to their favourable combinations of short periods, extended atmospheres and large planet-to-star radius ratios -- all of which helps to maximise the achievable signal-to-noise of what are very challenging measurements. The extreme temperatures experienced by the most highly irradiated hot-Jupiters may also lead to a simplification of the atmospheric chemistry on these planets, due to the thermal dissociation of a large fraction of molecular species into their constituent elements -- aiding in the interpretation of their atmospheric observations \citep{2018ApJ...866...27L,2018ApJ...863..183K}. The technique of transmission spectroscopy, which measures the variations in the effective size of the planet as a function of wavelength, is an enormously successful method for investigating the chemical composition and physical structure of a planet’s atmosphere \citep{2001ApJ...553.1006B,2000ApJ...537..916S}. Transmission spectroscopy observations have provided many high-significance detections of atomic and molecular species for a range of exoplanets, including detections of Na, K and H$_{2}$O both from the ground and using space-based instrumentation \citep[e.g.][]{2002ApJ...568..377C,2015MNRAS.446.2428S,2013ApJ...774...95D,2014ApJ...793L..27K,2014AJ....147..161S,2016ApJ...832..191N,2018Natur.557..526N,2016ApJ...822L...4E,2018AJ....156..283E,2018Natur.557...68S,2020MNRAS.494.5449C}. However, other observations have resulted in partially muted or even completely featureless spectra due to the influence of clouds and/or hazes \citep[e.g.][]{2011MNRAS.416.1443S,2016Natur.529...59S,2011ApJ...743...92B,2013MNRAS.432.2917P,2013MNRAS.436.2974G}.

A number of recent studies have also focused on investigating the temperature structures of highly irradiated exoplanets and the presence of vertical thermal inversions \citep[e.g.][]{2010ApJ...720.1569K,2011ApJ...743..191M,2015ApJ...813...47M,2017AJ....154..158B,2018A&A...617A.110P,2018ApJ...866...27L,2019MNRAS.485.5817G}. These inversion layers occur high in the planet’s atmosphere where the temperature is increasing with altitude and are present in the atmospheres of almost all of the Solar System planets \citep{1969ApJ...157..925G,1974ApJ...193..481W,1974ApJ...187L..41R}. The Earth’s stratosphere for example is due to absorption of UV photons high in the atmosphere by ozone and hazes. For very highly irradiated exoplanets it has been predicted that the molecules titanium oxide (TiO) and/or vanadium oxide (VO) could provide both the required opacity to incoming optical radiation and the necessary abundances to be strong candidates for the drivers of these thermal inversions \citep{2003ApJ...594.1011H,2008ApJ...678.1419F}. Though thus far, and despite much effort, detections exist only for a limited number of planets \citep[e.g.][]{2017AJ....154..221N,2017Natur.549..238S}, and in most cases have been disputed \citep[e.g.][]{2020AJ....160...93H,2019MNRAS.482.2065E}. The presence of a thermal inversion can be inferred from planetary emission spectra by observing bands in emission rather than absorption, and this was first achieved for the hot Jupiter WASP-121b from observations of H$_{2}$O on the dayside \citep{2017Natur.548...58E}, and has since been verified using TESS optical phase curves and other observations \citep{2019arXiv190903000D,2020A&A...637A..36B}.  

WASP-121b is a highly irradiated ultra-hot Jupiter discovered by \citet{2016MNRAS.458.4025D}. It orbits a bright F6-type star (V = 10.4) and has a highly inflated atmosphere, both of which make it particularly amenable to transmission spectroscopy observations. It has a mass similar to that of Jupiter and a radius significantly larger  -- 1.18\,\textit{M}$_\mathrm{J}$ and 1.7\,\textit{R}$_\mathrm{J}$ respectively, and an equilibrium temperature above 2400\,K. WASP-121b is amongst the most extensively studied of the transiting planets with numerous observations having being acquired from the ground and from space and at both low- resolution and using high-resolution Doppler-resolved spectroscopy. At low-resolution, observations obtained with the \textit{Hubble Space Telescope} (HST) have resulted in detections of H$_{2}$O along with excess optical absorption which was tentatively attributed to TiO/VO \citep{2016ApJ...822L...4E}. Follow-up observations revealed further evidence for VO \citep{2018AJ....156..283E} along with an unknown blue absorber suggested to be attributed to the SH molecule. However, additional secondary eclipse measurements have not been able to confirm the previous VO detection \citep{2019MNRAS.488.2222M,2020MNRAS.496.1638M} and instead suggest H$^{-}$ emission. 

It has not yet been possible to definitively show that either VO or TiO are responsible for the thermal inversion in the atmosphere of WASP-121b. Additionally, \citet{2017AJ....154..158B} and \citet{2013A&A...558A..91P} both show how cold-trap processes can interfere with the circulation of TiO/VO and suppress the formation of inversions, and it has also been demonstrated \citep{2018A&A...617A.110P,2018ApJ...866...27L} that for the very hottest planets much of the TiO and VO will be thermally dissociated on the dayside and in these cases thermal inversions could instead be driven by NUV and optical absorption by gas phase metals such as Fe and Mg. Indeed, at high-resolution, the atmosphere of WASP-121b has been observed using both the HARPS \citep{2020A&A...635A.205B,2020MNRAS.494..363C} and UVES \citep{2020MNRAS.493.2215G,2020A&A...636A.117M} spectrographs, which has resulted in significant detections of the atomic metals FeI and NaI and absorption from H-alpha but non-detections of both TiO and VO. FeI in particular is a strong optical absorber and its presence could therefore provide the source of the heating required to produce the observed temperature inversion \citep{2020MNRAS.493.2215G,2020ApJ...894L..27P}. 

Even more recent observations at high-resolution have confirmed the previous metal detections and also uncovered evidence for multiple other atomic species including Mg, Na, Ca, Cr, Fe, Ni, V \citep{2020ApJ...897L...5B,2020A&A...641A.123H} and a very recent detection of Li using ESPRESSO on the VLT \citep{2020arXiv201101245B}. In the UV, SWIFT/UVOT observations revealed tentative evidence of metal ions high in the atmosphere \citep{2019A&A...623A..57S}, whilst HST/STIS observations detected strong absorption lines from iron and magnesium atoms \citep{2019AJ....158...91S} extending far higher than the optical and near-infrared features and demonstrating the existence of an extended and possibly escaping atmosphere. In the present study we report optical transmission spectroscopy observations of WASP-121b using the Gemini Multi-Object Spectrograph (GMOS). Our observations aim to search for evidence of absorption by TiO/VO or other metals in order to infer the main driver of the thermal inversion by resolving the shape of these molecular features.

This paper is structured as follows: we describe our observations and data reduction steps in Section 2 and detail our light curve analysis in Section 3; in Section 4 we describe our atmospheric retrieval approach and discuss our results in Section 5. Finally, we offer our conclusions in Section 6.

\section{GMOS Observations and Data Reduction}

\begin{figure*}
 \includegraphics[width=\textwidth]{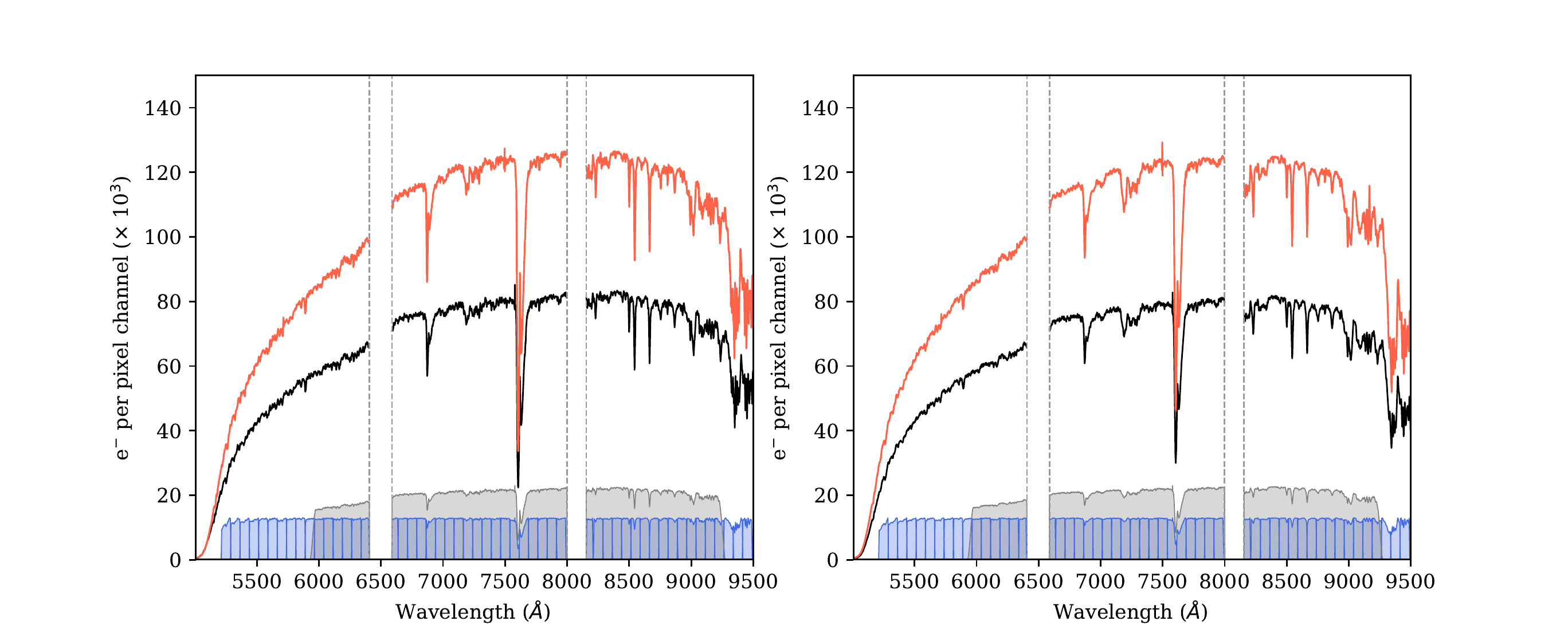}
 \caption{Example spectra of WASP-121 (black) and the comparison star (red) for Transit 1 (left) and Transit 2 (right). Coloured regions indicate the spectral bins used for extraction of the white light curves (grey) and spectroscopic light curves (blue). Dashed lines indicate the locations of the GMOS detector gaps.}
 \label{fig:examplespec}
\end{figure*}

Two transits of the ultra-hot Jupiter WASP-121b were observed on the nights 2017 January 4 and 2017 January 9 (hereafter Transit 1 and 2, respectively) using the 8-m Gemini-South telescope with the Gemini Multi-Object Spectrograph (GMOS, \citealt{2004PASP..116..425H}) as part of programme GS-2016B-Q-42 (PI: Gibson). GMOS consists of three 2k\,$\times$\,4k CCDs arranged side by side and separated by small detector gaps and has an imaging field-of-view of 5.5\,$\times$\,5.5 arcminutes squared. Both transits were observed using the R400 grism\,$+$\,OG515 filter with a central wavelength of 725 nm in 2\,$\times$\,2 binning mode covering the spectral range of 515\,$-$\,940 nm. To reduce readout time, we read out only three regions of interest (ROI)\footnote{To speed up the readout time only pixels containing flux from the slits were recorded into separate regions of interest (ROIs)} on the chip including the target and two comparison stars. For Transit 1 we obtained 580 science exposures over a total period of 340 minutes, whilst for Transit 2 we obtained a total of 505 science exposures covering a total period of 290 minutes. For both transits we used exposure times of 10 s with a readout time of 20 s. For Transit 1 the on-sky separation of the Moon from the target was $\approx$\,97 degrees with lunar illumination of $\approx$\,44\%. For Transit 2 the Moon was separated by $\approx$\,63 degrees with illumination of 93\%.

To enable differential spectroscopy we observed two comparison stars simultaneously with WASP-121 for each transit and used a custom mask with wide slit widths (40\,$\times$\,15 arcsec) in order to reduce the impact of differential slit losses. Only the brighter of our two comparison stars was used in the subsequent analysis since the other was found to be significantly fainter. The average FWHM for Transit 1 was found to be $\approx$\,4 pixels but varied between a maximum of $\approx$\,7 pixels and a minimum of $\approx$\,3 pixels, whilst for Transit 2 the average was $\approx$\,5 pixels, also varying between a maximum of $\approx$\,8 pixels and a minimum of $\approx$\,4 pixels. The seeing-limited spectral resolution was found to be R $\approx$ 690 – 1500 and R $\approx$ 600 – 1200 for Transit 1 and 2 respectively, and we observed the target from an airmass of 1.51 to 1.01 for Transit 1 and 1.25 to 1.01 for Transit 2. 

The standard GMOS pipeline contained in the Gemini IRAF\footnote{IRAF is distributed by the National Optical Astronomy Observatory, which is operated by the Association of Universities for Research in Astronomy (AURA) under cooperative agreement with the National Science Foundation}/PyRAF\footnote{PyRAF is a product of the Space Telescope Science Institute, which is operated by AURA for NASA} package was used to carry out the initial data reduction. We first converted the ROI images to standard GMOS format, and then proceeded with basic reduction procedures to de-bias and flat field the raw images using the calibration frames obtained at the beginning and end of each night. To extract the spectra we used a custom pipeline in IRAF/PyRAF and experimented with various aperture widths and background regions in order to minimise the residual scatter. For the final analysis we used an optimum aperture radius of 18 pixels and two background regions located 80 – 100 pixels either side of the spectral trace. The background contribution was estimated by taking the median value within these regions. Figure 1 shows example spectra  for WASP-121 and the comparison star for both transits obtained after spectral extraction. 

To align each of our spectra we use the x-shifts obtained from a process of cross-correlation using suitable absorption features in the stellar spectra after normalising the continua. We repeated this cross-correlation procedure for a number of different absorption features but found the results to be insensitive to the specific feature chosen, in addition, we did not notice any significant stretching of the spectra when compared to the width of our bins over the time-series, and we use the alignment obtained using the O$_{2}$ feature in our final analysis. 

To calibrate the wavelength scales, we first tried the wavelength solution derived from arc frames obtained with a calibration mask with much narrower 1 arcsec slit widths. However, inspection of our spectra revealed residual offsets between the target and comparison spectra and we instead constructed our solution by first identifying a set of well-resolved absorption lines in the mean spectrum, and then fitting a Gaussian to each of these lines to accurately retrieve the position of the line centres. Our solution is obtained by fitting these line centres using a Gaussian process (GP) with a standard squared exponential kernel (see Section 3.1 for a detailed description of our implementation of GPs) and we use the solution derived from this fit in our final analysis.

To produce the differential white light curves for each transit we bin the flux along the dispersion axis for the target and comparison star over the wide wavelength ranges shown in grey in Figure 1, and then divide the flux of the target star by the flux of the comparison star for the entire time-series to correct for variations in observing conditions and telluric effects. Multiple spectroscopic light curves are constructed in a similar way by integrating over each of the narrower bins shown in blue (avoiding areas which overlapped the detector gaps). We extracted 53 spectroscopic channels for both transits and the resulting differential white light curves and spectral light curves are shown in Figures 2, 3 and 4.

Our limb darkening coefficients and their associated uncertainties were obtained using the PyLDTk toolkit \citep{2015MNRAS.453.3821P}, which uses the PHOENIX models of \citet{2013A&A...553A...6H}. For this we used the stellar parameters for WASP-121 and the system parameters and uncertainties as given in \citep{2016MNRAS.458.4025D}. Finally, we also examined various diagnostic measurements, including the FWHM of the spectral trace and the shifts in the dispersion and spatial axes to check for correlations with the instrumental systematics \citep[e.g.][]{2001ApJ...553.1006B,2003nicm.rept....1G,2007A&A...476.1347P,2009ApJ...690L.114S,2010Natur.464.1161S,2012A&A...542A...4G,2013MNRAS.434.3252H,2016ApJ...832..191N}, though in this case no obvious correlations were found.

\section{Analysis}
\subsection{White Light Curve Analysis}

To model our white light curves we use a very similar procedure to that previously outlined in \citet{2020MNRAS.497.5155W}, which implements the methodology introduced in \citet{2012MNRAS.419.2683G}. In brief, we fit using a time-dependent GP\footnote{For the implementation of our Bayesian inference we made extensive use of the Python modules \textsc{GeePea} and \textsc{Infer} which are freely available from \url{https://github.com/nealegibson}} simultaneously with the analytic transit model of \citet{2002ApJ...580L.171M} with quadratic limb darkening. GPs have been successfully implemented in a large number of similar studies \citep[e.g.][]{2013MNRAS.428.3680G,2017Natur.548...58E,2018AJ....156..283E,2018Natur.557..526N} and have been shown to be extremely useful for modelling correlated noise in exoplanet time-series, whilst offering robust uncertainty estimates. Our joint posterior distribution is given by the multivariate Gaussian: 
\begin{equation}
p(\bmath f| \bmath t,\bphi,\btheta) = \mathcal{N} \left (T(\bmath t,\bphi) , \boldsymbol{\Sigma} (\bmath t,\btheta) \right).
\end{equation}
where $\bmath t$ and $\bmath f$ are the vectors of time and flux measurements respectively, \textit{T} is the analytic transit mean function with parameters $\bphi$ and $\boldsymbol{\Sigma}$ is the covariance matrix with hyperparameters $\btheta$. Correlations between data points are described using the covariance matrix which is calculated using a kernel function with parameters $\boldsymbol{\theta}$ (for a more detailed review of GPs and kernels see \citealt{10.5555/1162254}). In this work we use the Mat\'ern 3/2 kernel which can be viewed as a less smooth version of the more commonly used squared exponential kernel and which is defined as:
\begin{equation}
    k({t}_n, {t}_m | \btheta) = \xi^2 \left( 1+{\sqrt{3}\,\eta\,\Delta t} \right) \exp \left( -{\sqrt{3}\,\eta\,\Delta t}\right) + \delta_{nm}\sigma^2,
\end{equation}
where $\xi$ is the height scale, $\Delta$\,t is the time difference of observations, $\eta$ is the inverse length scale, $\delta_{nm}$ is the Kronecker delta and $\sigma$ is the white noise term which is identical for each data point. As a check we repeated our analysis using a squared exponential kernel but found that our results were unaffected. Since no obvious correlations were observed between the form of the systematics in the light curves and the diagnostic measurements, we proceeded by using time as our only input variable, which has the added advantage of requiring fewer assumptions. We obtain the posterior probability distributions of the parameters of interest by first specifying prior distributions for the hyperparameters of the kernel function and then multiplying by the marginal log-likelihood.

Our GP mean function assumes a circular orbit with period fixed to that reported by \citet{2016MNRAS.458.4025D}. For each white light curve we allow the mid-transit time (\textit{T}$_\mathrm{c}$) and linear baseline parameters (\textit{f}$_{\mathrm{oot}}$, \textit{T}$_{\mathrm{grad}}$) to vary as free parameters and fixed the system scale (\textit{a}/\textit{R}$_\mathrm{\star}$) and impact parameter \textit{b} to the values previously constrained in \citet{2018AJ....156..283E}. For the planet-to-star radius ratio ($\rho$ = \textit{R}$_\mathrm{p}$/\textit{R}$_\mathrm{\star}$) we use a Gaussian prior centered on the  reported value in \citet{2018AJ....156..283E}. Constraining the white light curve parameters to previously reported values allows us to more easily compare the respective results and should help to improve the accuracy of our systematics models.

To account for the effects of stellar limb darkening, we used a quadratic limb darkening law \citep{2000A&A...363.1081C} and placed Gaussian priors on the quadratic coefficients \textit{c}$_\mathrm{1}$ and \textit{c}$_\mathrm{2}$ with a mean and standard deviation determined using PyLDTk. We also tried repeating our analysis having both fixed the limb darkening parameters to their best-fit values and leaving the parameters completely free in the fits, but found that this variation in treatment did not affect our results (see Section 5 where we discuss the impact of our limb darkening treatment further). We summarise the assumed parameter values for the white light curves in Table 1. Similar to previous studies \citep[e.g.][]{2017MNRAS.467.4591G,2017Natur.548...58E}, we fit for log $\eta$ and log $\xi$ with uniform priors and constrain the length scales to lie between the cadence and twice the total duration of our observations. 

We performed a joint fit to both individual transits, allowing for only a single planet-to-star radius ratio and limb darkening coefficient pair, whilst allowing the hyperparameters to vary separately for each transit. To check the validity of our assumed parameter values, we also carried out independent fits to both of our individual white light curves finding the values to be consistent within 1 sigma to those of \citet{2018AJ....156..283E} for both transits, though the measured planet-to-star radius ratio was found to be higher for one transit and slightly lower for the other (discussed further in Section 5). This ultimately results in an offset in the mean levels of the individual transmission spectra, though it does not affect their relative values and we find that both transits produce spectra with a consistent shape. We further tested this by again fitting each transit independently, but this time using the same priors as for our joint fit (shown in Figure 5). We confirm that the shape of the individual spectra are not affected by imposing these priors and find that both transits produce spectra which are in excellent agreement, though with quite different average uncertainties. We computed the reduced $\chi^2$ between the two spectra to be 0.7 by taking the difference between each pair of points and adding the uncertainties in quadrature \footnote{where the difference between pairs of data points will follow a normal distribution $\Delta x \sim \mathcal{N}(0,\sigma_{x1}^2+\sigma_{x2}^2)$, hence distributed according to $\chi^2$}. The overall mean level of the transmission spectrum can also be influenced by inaccuracies in the common-mode correction, though this similarly should not affect the relative values, but can potentially lead to offsets, especially when including multiple datasets from different instruments, and needs to be carefully considered when interpreting results. The individual transmission spectra for Transits 1 and 2 are shown in Figure 5 and the results from the joint fits are shown in Figure 6. We discuss these results further in Section 5.

To find the best-fitting models for each of our white light curves we began by using a differential evolution algorithm to optimise the log-likelihood with respect to the transit and kernel parameters using the values from \citet{2016MNRAS.458.4025D} and \citet{2018AJ....156..283E} as the initial points. Next, we used a Nelder-Mead simplex algorithm \citep{10.1093/comjnl/7.4.308} to fine-tune these estimated values and, finally, we ran a Markov-Chain Monte-Carlo (MCMC) method to marginalise our posterior distributions. Four independent chains with length 80,000 were initiated for each of our light curves with the first 40$\%$ of samples in the chain being considered as burn-in and discarded. We used the Gelman-Rubin statistic to confirm mutual convergence for each free parameter. We show the best-fit white light curve models and derived systematics models and residuals for both transits in Figure 2.

\begin{figure}
 \includegraphics[width=1.07\columnwidth]{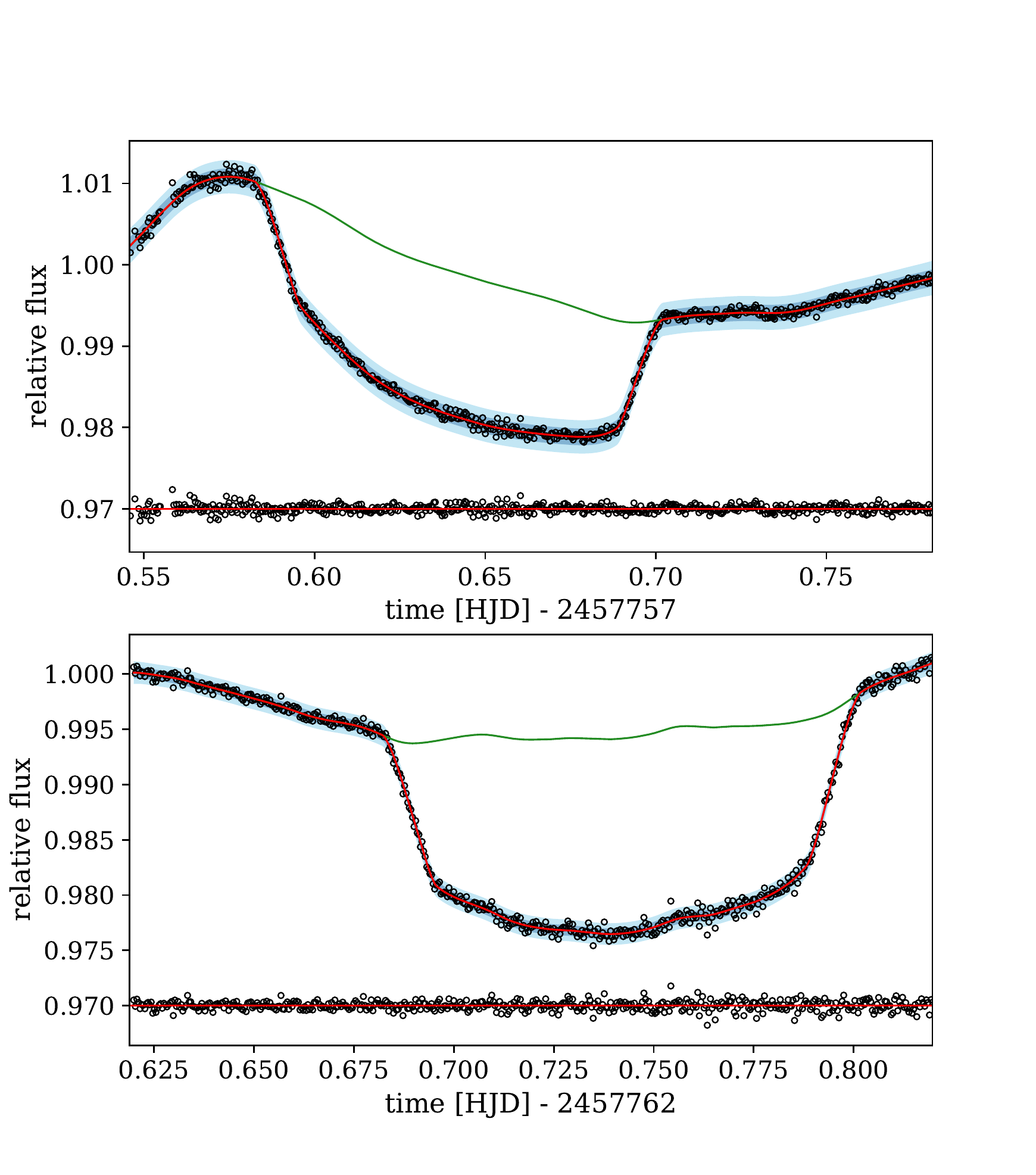}
 \caption{White light curves of WASP-121b obtained with the R400 grism + OG515 filter for Transit 1 (top) and Transit 2 (bottom). The red line shows the best fit model with blue shading indicating plus/minus two standard deviations. The green line shows the systematics model derived from the GP fit. Residuals are indicated below the light curves.}
 \label{fig:wlc}
\end{figure}

\begin{table}
 \caption{Assumed transit parameter values used in the fitting of the white light curves. The orbital period, system scale and impact parameter were held fixed and Gaussian priors were placed on the following parameters with the mean and standard deviations given below.}
 \label{tab1}
 \begin{tabular}{ll}
  \hline
  Parameter & Value\\
  \hline
  \textit{P} & 1.2749255 days (fixed)\\[2pt] 
  \textit{a}/\textit{R}$_\mathrm{\star}$ & 3.86  (fixed)\\[2pt]
  \textit{b} & 0.06 (fixed)\\[2pt]
  \textit{$\rho$} & 0.1219 $\pm$ 0.0005\\[2pt]
  \textit{c}$_\mathrm{1}$ & 0.395 $\pm$ 0.003\\[2pt]
  \textit{c}$_\mathrm{2}$ & 0.141 $\pm$ 0.004\\[2pt]
  \hline
 \end{tabular}
\end{table}

\subsection{Spectroscopic Light Curve Analysis}

\begin{figure*}
 \includegraphics[ width=\textwidth]{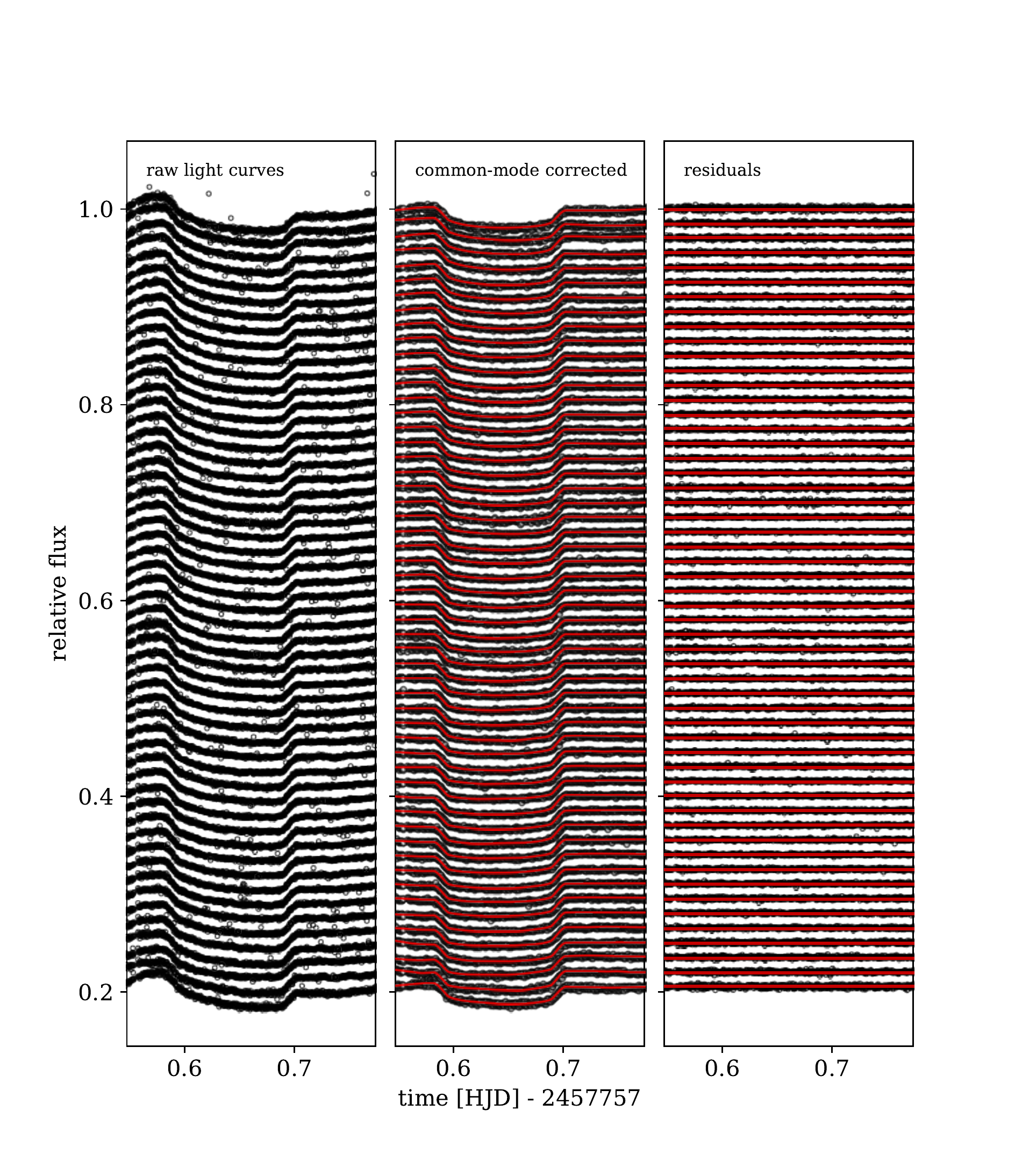}
 \caption{Spectral light curves for the R400 grism + OG515 filter for Transit 1 corresponding to the spectral channels shown in blue in Figure 1 (left). The left panel shows the raw light curves before correction. The middle panel shows the light curves with best fit GP model after the common-mode correction. The right panel shows the residuals from the best-fit model.}
 \label{fig:slc1}
\end{figure*}

To construct our spectroscopic light curves, we began by binning over the narrow wavelength regions shown in blue in Figure 1. For each transit, we extracted 53 individual light curves using a uniform bin width of 75 Å whilst avoiding the GMOS detector gaps. We show the resulting raw light curves in the left panels of Figures 3 and 4. The raw light curves for both transits show significant systematics which are largely independent of wavelength. This is a similar situation to that encountered for the FORS2 dataset in \citet{2020MNRAS.497.5155W} and we proceeded in much the same way by dividing each of the spectroscopic light curves by the common-mode corrections derived from the white light curves. Additionally, to remove any remaining high-frequency systematics, we also subtract the residuals from the white light curves fits. The net result of this procedure is to provide significant improvements to the precision of our transmission spectrum whilst preserving the relative values of the planet-to-star radius ratios. 

After applying the common-mode correction described above we find the best-fitting model for each spectroscopic light curve using the same process and the same systematics model as described in Section 3.1. The main difference is that we fix the mid-transit time to the best-fit value from the white light curve analysis, whilst allowing the planet-to-star radius ratio, limb darkening parameters, linear baseline parameters and kernel hyperparameters to vary for each fit. As for the white light curves, we perform a joint analysis allowing for only a single planet-to-star radius ratio and limb darkening coefficient pair for each individual wavelength channel, but allow the hyperparameters to vary separately for each transit. To add another layer of flexibility to our model we use wide Gaussian priors with a standard deviation of 0.1 for the limb darkening coefficients, with a mean given by the best fit values determined using PyLDTk. We again used the same differential evolution algorithm to optimise the log-likelihood as for the white light curves, and fine-tuned using a Nelder-Mead simplex algorithm. Outliers deviating more than 4\,$\sigma$ from the predictive distribution were removed for each fit (this procedure typically clipped 1-2 points per light curve) and we used the same MCMC method described above to marginalise our posterior distributions. 
The best-fit common-mode corrected GP models for the spectroscopic light curves are shown in the middle panels of Figures 3 and 4 and we list the measured planet-to-star radius ratios and uncertainties in Table 2. Our GMOS transmission spectrum for WASP-121b is shown in Figure 6 and the results are discussed further in Section 5. 

\begin{figure*}
 \includegraphics[width=\textwidth]{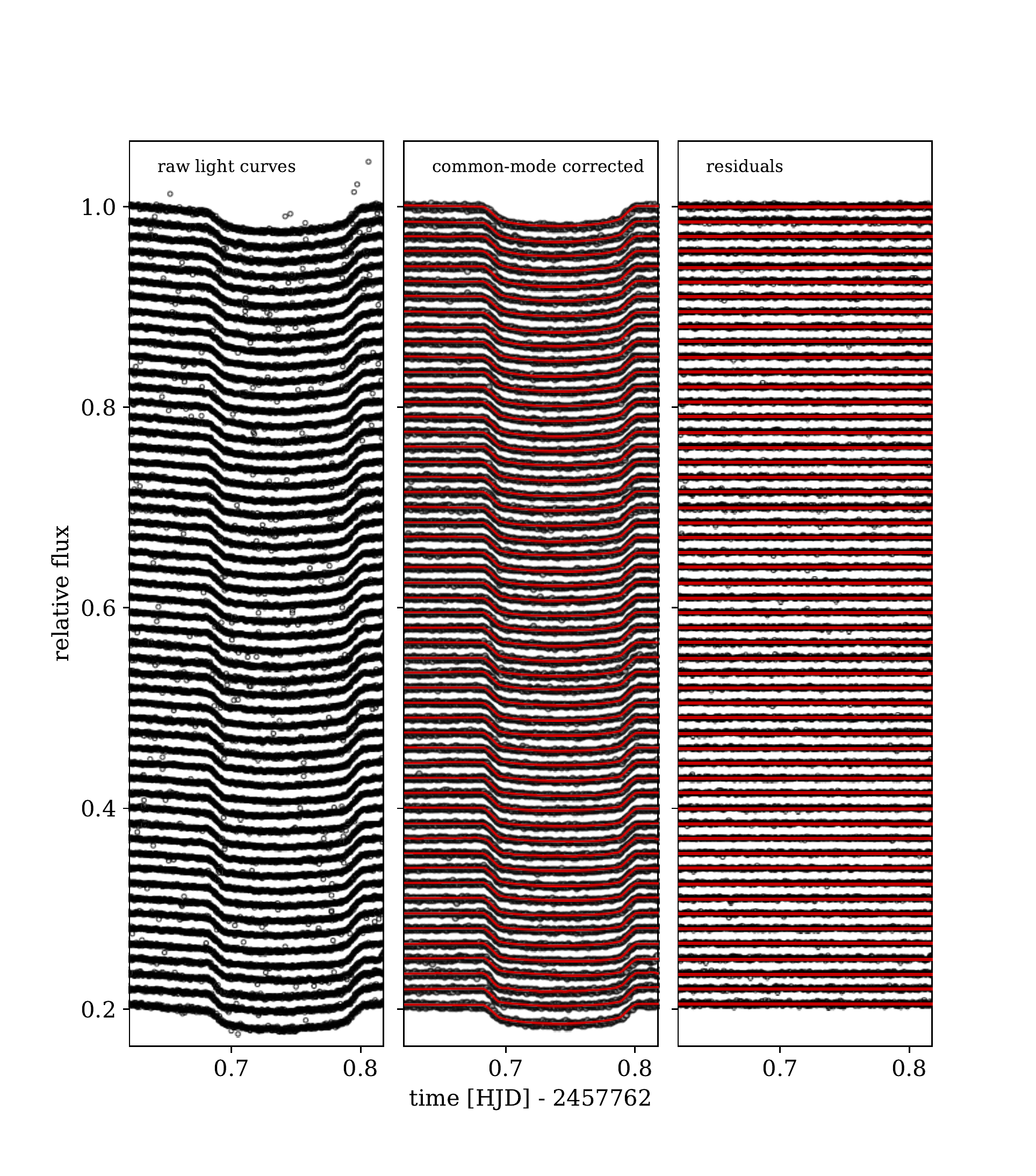}
 \caption{Same as Figure 3 but for Transit 2. The left panel shows the raw light curves before correction. The middle panel shows the light curves with best fit GP model after the common-mode correction. The right panel shows the residuals from the best-fit model.}
 \label{fig:slc2}
\end{figure*}

\subsection{Modelling Systematics with Student's-T Processes}

To investigate the robustness of the analysis outlined above, we performed some additional modelling using a Student's-T Process (STP), implemented with a modified version of our GeePea code. STPs have already been successfully applied to Bayesian optimisation problems and aerospace design \citep[e.g.][]{Shah2013BayesianOU,doi:10.2514/6.2018-1659} and have been shown to come at no additional computational cost over GPs \citep{pmlr-v33-shah14} but, to our knowledge, have not yet been applied to the analysis of systematics in transit light curves.

Whilst the basis of a GP is the multivariate Gaussian distribution, the basis of a STP is instead the multivariate Student's-T distribution \citep{book} and therefore, similarly to a GP, a STP defines a prior over functions. A STP differs however from a GP in two ways: firstly, a Student's-T distribution can have a higher kurtosis than the corresponding Gaussian distribution, and therefore a STP prior can assign higher probability to extreme outliers. Secondly, with a GP, the conditioned posterior variance (excluding white noise) depends exclusively on the location of the input variables, whilst with a STP the posterior variance will also depend explicitly on the observed values. To account for this extra complexity, a STP has an additional parameter, $\nu$, which describes the degrees of freedom of the distribution. As the degrees of freedom approaches infinity, the multivariate Student’s-T distribution converges to the corresponding multivariate Gaussian distribution. Thus, a STP can be considered as a generalisation of a GP with an additional parameter. Following the derivation outlined in \citet{doi:10.2514/6.2018-1659}, the pdf of a STP is given by:
\begin{equation}
\begin{split}
\mathcal{T}(\mu, \bmath \Sigma, \nu) & = \frac{\Gamma\left(\left(\nu + d \right)/2\right)}{\Gamma\left(\nu/2\right)\nu^{d/2}\pi^{d/2}|\bmath \Sigma|^{1/2}} \\
& \times \left( 1+\frac{1}{\nu}\left(y - \mu \right)^T\bmath \Sigma^{-1}\left(y - \mu \right)\right)^{-\left(\nu + d \right)/2},
\end{split}
\end{equation}

where $\mu$ and $\nu$ are the mean and degrees of freedom of the distribution respectively and $d$ is the dimension. $\bmath \Sigma$ is a symmetric positive definite shape parameter which is related to the covariance matrix by:
\begin{equation}
    E[\left(y - \mu \right)^T\left(y - \mu\right)] = \frac{\nu}{\nu - 2}\bmath \Sigma.
\end{equation}

 Using the notation common in the GP literature and given a set of observations $\bmath D$ = \{($x_1$,$y_1$), ($x_2$,$y_2$), ($x_3$,$y_3$),...\}, the posterior mean and covariance will be given by:
\begin{equation}
    \hat{\mu} = \bmath K_{\star} \bmath K^{-1} \bmath y,
\end{equation}
and:
\begin{equation}
    \hat{K} = \frac{\nu +\bmath y^T \bmath K^{-1}\bmath y-2 }{\nu + |\bmath D| - 2} \left(\bmath K_{\star\star}- \bmath K_{\star} \bmath K^{-1} \bmath K_{\star}^T\right).
\end{equation}

These two equations are the same as that which would be obtained for the equivalent GP, except for an additional term modifying the covariance, which can be seen as a corrective factor which depends explicitly on the observed values. In short, if the observed values are consistent with a GP then the posterior covariance of the STP will be roughly equal to that of a GP. Conversely, if the variation in the observed values is greater or smaller than expected then the STP posterior covariance will be larger or smaller respectively. In other words, for the same kernel, the posterior means will be identical, but the variance will differ, depending on the observed values.

For our implementation of STPs we used the same Mat\'ern 3/2 kernel and hyperparameter priors as for our GP analysis described in Sections 3.1 and 3.2. For the degrees of freedom parameter we set a prior condition of $\nu$ > 4 to ensure that the distribution has both finite variance and finite kurtosis. We then carry out the same procedure as before to perform a joint fit to both transits. We compare the results from the STP analysis with those from our GP analysis in Figure A1 in the appendix. As can be seen from the figure our results are almost identical across the full spectrum, with a mean uncertainty that is inflated by only a few percent and a posterior mean for $\nu$ that was typically a very large number. Our results therefore verify that the observed data can indeed be well described by a GP in this case, and gives us greater confidence in our estimated uncertainties. 

For the analysis as implemented here we should acknowledge that the value for $\nu$ is common for both the stochastic component, and for the noise. Ideally, these should be treated independently with separate instances of $\nu$ for each. However, the fact that we recover almost identical results to our GP analysis is reassuring. We will return to examine this potential shortcoming in greater detail in future work. A more detailed discussion on the comparisons between STPs and GPs and their respective strengths and weaknesses is beyond the scope of the present study.

\begin{table}
 \caption{Transmission spectrum for WASP-121b recovered from the GMOS low-resolution spectroscopic light curves. }
 \label{tab2}
 \begin{tabular}{lcc}
  \hline
  Wavelength  & Radius Ratio & Limb Darkening\\
  Centre [Range] ({\AA}) & \textit{R}$_\mathrm{p}$/\textit{R}$_\mathrm{\star}$ & \textit{c}$_\mathrm{1}$ \hspace{0.5cm} \textit{c}$_\mathrm{2}$\\
  \hline
  5250 [5213-5288] & 0.12413 $\pm$ 0.00148 & 0.584\hspace{0.3cm}0.113\\
  5325 [5288-5363] & 0.12501 $\pm$ 0.00136 & 0.576\hspace{0.3cm}0.115\\
  5400 [5363-5438] & 0.12411 $\pm$ 0.00079 & 0.561\hspace{0.3cm}0.117 \\
  5475 [5438-5513] & 0.12490 $\pm$ 0.00065 & 0.557\hspace{0.3cm}0.120\\
  5550 [5513-5588] & 0.12512 $\pm$ 0.00050 & 0.549\hspace{0.3cm}0.124\\
  5625 [5588-5663] & 0.12527 $\pm$ 0.00122 & 0.537\hspace{0.3cm}0.126\\
  5700 [5663-5738] & 0.12440 $\pm$ 0.00085 & 0.527\hspace{0.3cm}0.130\\
  5775 [5738-5813] & 0.12341 $\pm$ 0.00137 & 0.523\hspace{0.3cm}0.133\\
  5850 [5813-5888] & 0.12508 $\pm$ 0.00090 & 0.509\hspace{0.3cm}0.137\\
  5925 [5888-5963] & 0.12485 $\pm$ 0.00086 & 0.508\hspace{0.3cm}0.134\\
  6000 [5963-6038] & 0.12397 $\pm$ 0.00117 & 0.498\hspace{0.3cm}0.136\\
  6075 [6038-6113] & 0.12373 $\pm$ 0.00045 & 0.493\hspace{0.3cm}0.135\\
  6150 [6113-6188] & 0.12441 $\pm$ 0.00058 & 0.485\hspace{0.3cm}0.131\\
  6225 [6188-6263] & 0.12482 $\pm$ 0.00045 & 0.474\hspace{0.3cm}0.135 \\
  6300 [6263-6338] & 0.12417 $\pm$ 0.00099 & 0.473\hspace{0.3cm}0.136\\
  6370 [6333-6408] & 0.12345 $\pm$ 0.00090 & 0.464\hspace{0.3cm}0.138 \\
  6610 [6580-6640] & 0.12264 $\pm$ 0.00048 & 0.424\hspace{0.3cm}0.155 \\
  6675 [6650-6700] & 0.12305 $\pm$ 0.00086 & 0.438\hspace{0.3cm}0.142 \\
  6750 [6713-6788] & 0.12421 $\pm$ 0.00086 & 0.433\hspace{0.3cm}0.141 \\
  6825 [6788-6863] & 0.12348 $\pm$ 0.00049 & 0.429\hspace{0.3cm}0.138 \\
  6900 [6863-6938] & 0.12367 $\pm$ 0.00104 & 0.424\hspace{0.3cm}0.144 \\
  6975 [6938-7013] & 0.12434 $\pm$ 0.00054 & 0.418\hspace{0.3cm}0.144 \\
  7050 [7013-7088] & 0.12337 $\pm$ 0.00054 & 0.414\hspace{0.3cm}0.139 \\
  7125 [7088-7163] & 0.12358 $\pm$ 0.00066 & 0.405\hspace{0.3cm}0.141 \\
  7200 [7163-7238] & 0.12317 $\pm$ 0.00037 & 0.405\hspace{0.3cm}0.140 \\
  7275 [7238-7313] & 0.12223 $\pm$ 0.00051 & 0.398\hspace{0.3cm}0.140 \\
  7350 [7313-7388] & 0.12193 $\pm$ 0.00076 & 0.395\hspace{0.3cm}0.140 \\
  7425 [7388-7463] & 0.12279 $\pm$ 0.00037 & 0.388\hspace{0.3cm}0.141 \\
  7500 [7463-7538] & 0.12311 $\pm$ 0.00063 & 0.384\hspace{0.3cm}0.142 \\
  7575 [7538-7613] & 0.12261 $\pm$ 0.00044 & 0.379\hspace{0.3cm}0.141 \\
  7650 [7613-7688] & 0.12292 $\pm$ 0.00044 & 0.373\hspace{0.3cm}0.141 \\
  7725 [7688-7763] & 0.12158 $\pm$ 0.00043 & 0.369\hspace{0.3cm}0.142 \\
  7800 [7763-7838] & 0.12129 $\pm$ 0.00040 & 0.366\hspace{0.3cm}0.141 \\
  7875 [7838-7913] & 0.12154 $\pm$ 0.00042 & 0.362\hspace{0.3cm}0.142 \\
  7950 [7913-7988] & 0.12218 $\pm$ 0.00042 & 0.358\hspace{0.3cm}0.140 \\
  8180 [8143-8218] & 0.12234 $\pm$ 0.00050 & 0.338\hspace{0.3cm}0.144 \\
  8250 [8225-8275] & 0.12160 $\pm$ 0.00055 & 0.330\hspace{0.3cm}0.147 \\
  8325 [8288-8363] & 0.12362 $\pm$ 0.00112 & 0.332\hspace{0.3cm}0.142 \\
  8400 [8363-8438] & 0.12230 $\pm$ 0.00046 & 0.317\hspace{0.3cm}0.150 \\
  8475 [8438-8513] & 0.12188 $\pm$ 0.00073 & 0.312\hspace{0.3cm}0.147 \\
  8550 [8513-8588] & 0.12384 $\pm$ 0.00043 & 0.307\hspace{0.3cm}0.147 \\
  8625 [8588-8663] & 0.12135 $\pm$ 0.00139 & 0.307\hspace{0.3cm}0.151 \\
  8700 [8663-8738] & 0.12184 $\pm$ 0.00075 & 0.306\hspace{0.3cm}0.148 \\
  8775 [8738-8813] & 0.12108 $\pm$ 0.00042 & 0.304\hspace{0.3cm}0.150 \\
  8850 [8813-8888] & 0.12170 $\pm$ 0.00088 & 0.288\hspace{0.3cm}0.157 \\
  8925 [8888-8963] & 0.12294 $\pm$ 0.00136 & 0.329\hspace{0.3cm}0.144 \\
  9000 [8963-9038] & 0.12266 $\pm$ 0.00048 & 0.279\hspace{0.3cm}0.158 \\
  9075 [9038-9113] & 0.12133 $\pm$ 0.00075 & 0.320\hspace{0.3cm}0.140 \\
  9150 [9113-9188] & 0.12235 $\pm$ 0.00046 & 0.331\hspace{0.3cm}0.139 \\
  9225 [9188-9263] & 0.12248 $\pm$ 0.00117 & 0.258\hspace{0.3cm}0.164 \\
  9300 [9263-9338] & 0.11829 $\pm$ 0.00178 & 0.325\hspace{0.3cm}0.138 \\
  9375 [9338-9413] & 0.12271 $\pm$ 0.00077 & 0.328\hspace{0.3cm}0.132 \\
  9450 [9413-9488] & 0.12228 $\pm$ 0.00138 & 0.320\hspace{0.3cm}0.137 \\
  \hline
 \end{tabular}
\end{table}

\begin{figure*}
 \includegraphics[width=\textwidth]{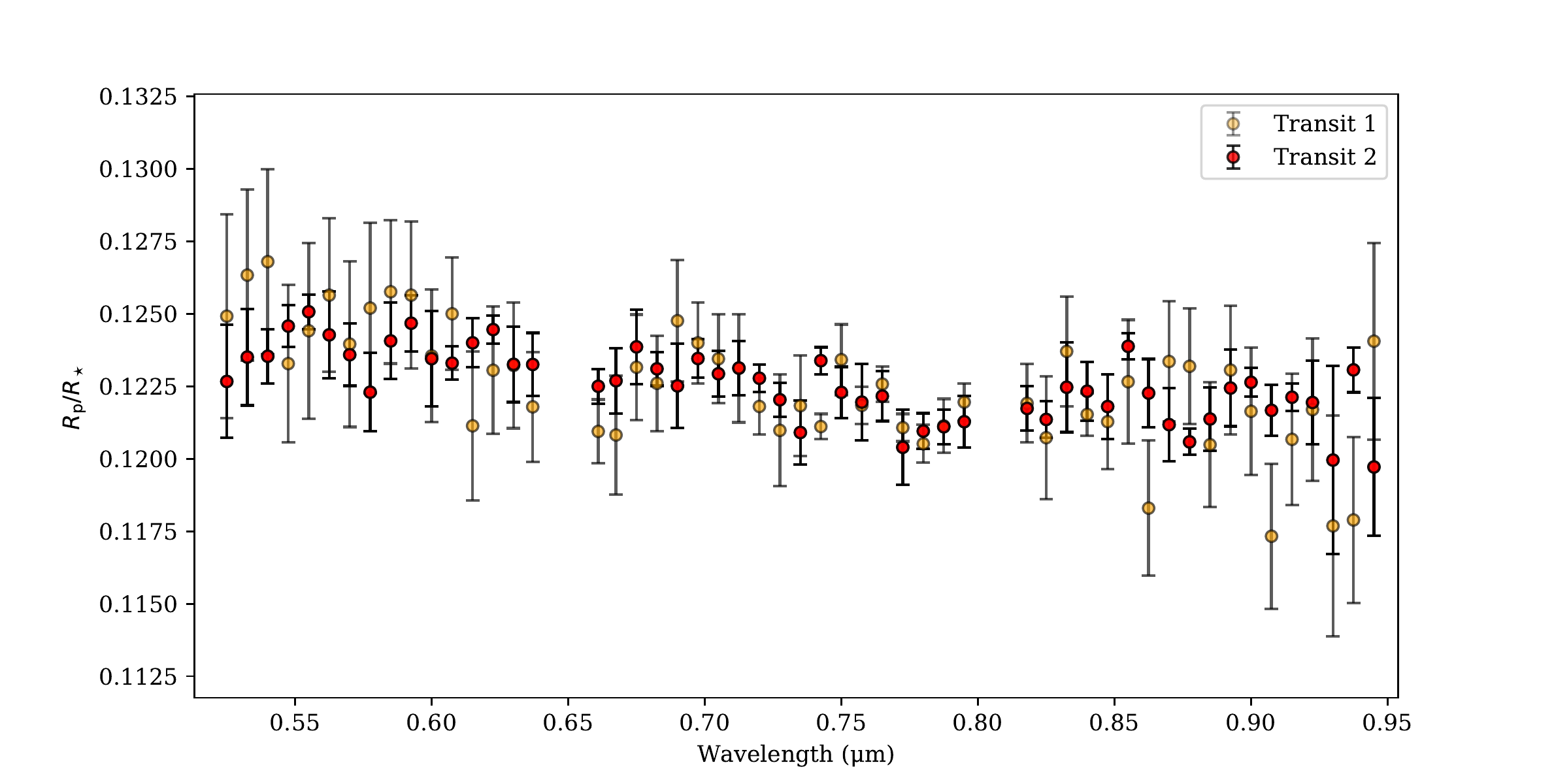}
 \caption{Individual Gemini/GMOS transmission spectra of WASP-121b for both transits. The orange points are the results of an independent analysis of the Transit 1 light curves using the same prior as for our joint analysis. The red points show the same but for Transit 2. The results from the joint fits are shown in Figure 6 below.}
 \label{fig:fors_spec}
\end{figure*}

\begin{figure*}
 \includegraphics[width=\textwidth]{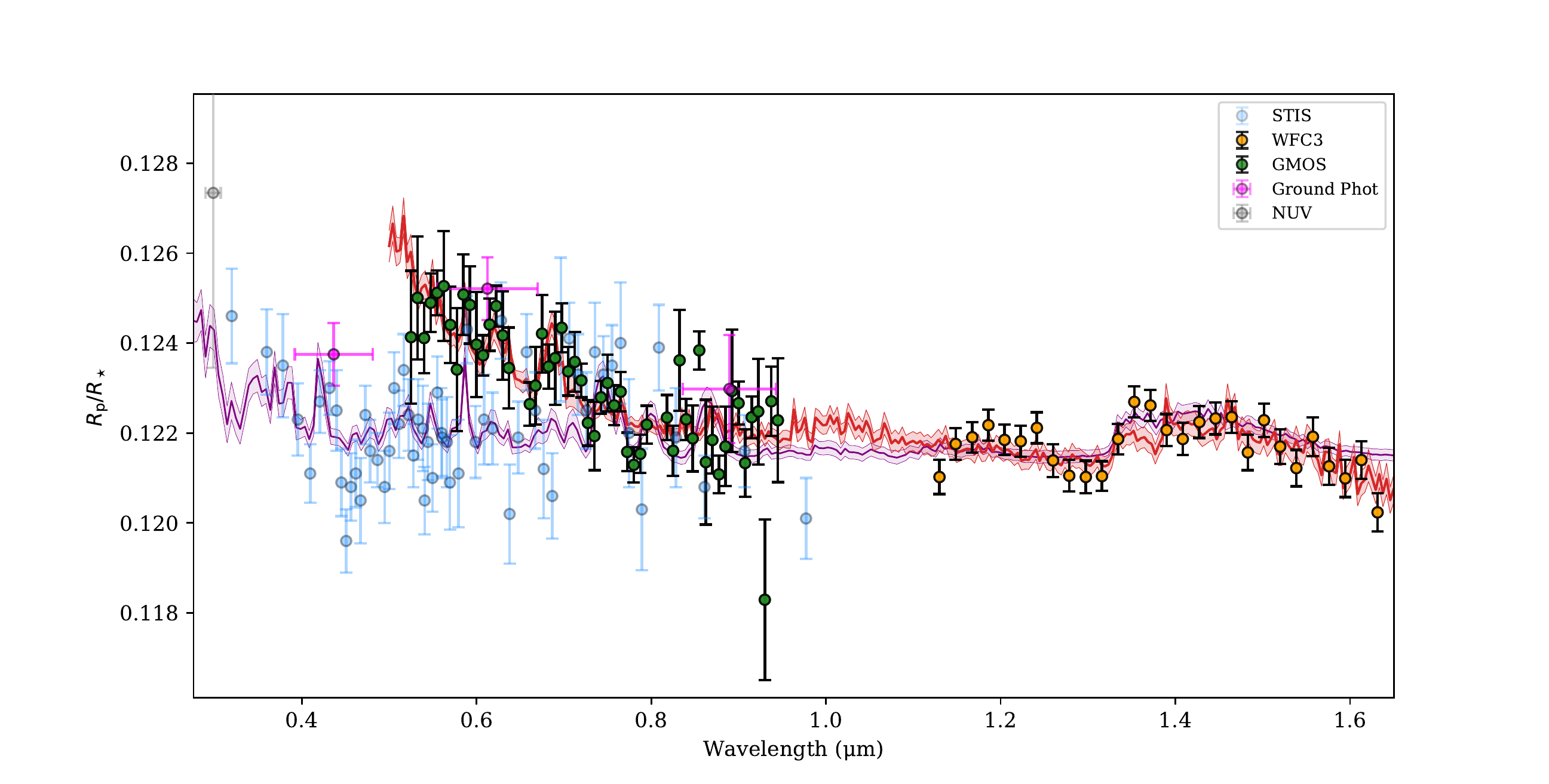}
 \caption{Combined optical-infrared transmission spectrum of WASP-121b obtained from Gemini/GMOS and WFC3 observations. The green points are the results from the joint fits to the GMOS light curves. The orange points are the WFC3 results from \citet{2016ApJ...822L...4E} which have been slightly shifted to match the best-fit offset found in the retrieval. The light blue points show the STIS results from \citet{2018AJ....156..283E}, magenta points show the results of a re-analysis of the ground-based photometry measurements from \citet{2016MNRAS.458.4025D} and the grey point on the far left indicates the longest wavelength NUV measurement from \citet{2019AJ....158...91S}. The red line shows the best-fit from our GMOS/WFC3 retrieval analysis using PETRA along with 1\,$\sigma$ significance contours (light red). We also show the best-fit model from the STIS/WFC3 retrieval (purple) described in \citet{2020arXiv201110626L} for comparison.}
\end{figure*}

\section{Atmospheric Modelling with PETRA}

To obtain constraints on the vertical temperature structure and terminator composition of WASP-121b, we performed an atmospheric retrieval using an adaptation of the retrieval code PETRA (PHOENIX ExoplaneT Retrieval Algorithm; \citealt{2020AJ....159..289L}). In addition to the optical GMOS data presented in Table 2, our retrieval also incorporates the near-IR HST/WFC3 data of \citet{2016ApJ...822L...4E}. Atmospheric retrieval codes typically comprise a forward model which is used to generate the predicted atmospheric spectrum and include a statistically robust inference method for parameter estimation. PETRA incorporates a modified version of the well-tested and self-consistent atmosphere model PHOENIX as its forward model, which has been widely used to study the atmospheres of both stellar and sub-stellar atmospheres \citep{1997ApJ...483..390H,1999ApJ...512..377H,2011ASPC..448...91A,2001ApJ...556..885B,2011ApJ...733...65B,2018ApJ...866...27L,2019ApJ...876...69L}. 
 
The model assumes hydrostatic equilibrium and a plane parallel atmosphere and solves for radiative transfer in a line-by-line manner for transmission geometry. PHOENIX incorporates an expansive line list database including CIA, bound-free and free-free opacities. We include the prominent absorbers expected for a hydrogen-rich atmosphere in the observed spectral range: H$_{2}$O, Na, K, TiO, VO, HCN and other metal oxides, hydrides, and atomic metal opacity. Chemical equilibrium is calculated with PHOENIX’s Astrophysical Chemical Equilibrium Solver (ACES) to calculate the equation of state for each species with elemental abundances scaled uniformly by a free-parameter for the bulk metallicity. We treat H$_{2}$O, TiO, and VO abundances as free and independent parameters. Line profiles are calculated using Voigt profiles. Details of the sources of the various opacity cross-sections can be found in \citet{2018ApJ...866...27L} and \citet{2019ApJ...876...69L}. 
 
 For the p-T profile we make use of the three-channel Eddington approximation outlined in \citet{2014A&A...562A.133P} as implemented in \citet{2013ApJ...775..137L}. This approach has five free parameters which describe the Planck mean thermal IR opacity, two independent downwelling visible channels of radiation, the partition of the flux between the two visible streams and a catch-all term which describes the albedo, emissivity, and day–night redistribution. Our model also considers the contributions from cloud/haze coverage as described in \citet{2017MNRAS.469.1979M}, with parameters $P_{\rm cloud}$ (the cloud-deck altitude in bars), a parameter which describes the scattering enhancement factor and a parameter which describes the scattering slope. For a full description of the retrieval framework see \citet{2020AJ....159..289L}.
 
 We use a Differential Evolution Markov Chain (DEMC) algorithm \citep{cite-key} with “snooker” updates \citep{cite-key2} as a statistical sampling algorithm to obtain the posterior distributions of the forward model parameters, checking for convergence using the Gelman-Rubin statistic. Lastly, we also include an offset between the GMOS/HST datasets as another free parameter to account for potential discrepancies. This is particularly important given that the \citet{2016ApJ...822L...4E} analysis adopted slightly different values for the transit shape parameters to those adopted in \citet{2018AJ....156..283E}. We used wide uniform priors for the temperature parameters, the H$_{2}$O, TiO, and VO abundances (between 10$^{-12}$ and 10$^{-1}$) and [Fe/H] (between 10$^{-1}$ and 10$^{4}$). For our final parameter and uncertainty estimates, we also inflated the error bars by $\sim$ 13$\%$ so that our fits were approaching a chi-square of 1.0.
 
 Our retrieval results are summarised in Table 3. We find little evidence for either TiO or VO with 1\,$\sigma$ upper limits on the best-fit absolute terminator abundances of log(\textit{X}$_\mathrm{TiO}$) < --9.60 and log(\textit{X}$_\mathrm{VO}$) < --10.25 respectively. We also constrain the H$_2$O abundance to log(\textit{X}$_\mathrm{H_2O}$) = --5.05$^\mathrm{+0.19}_\mathrm{-0.18}$ and find a best-fit mean temperature of 2664$^\mathrm{+120}_\mathrm{-100}$\,K. In contrast to the results in \citet{2018AJ....156..283E}, our retrieval favours a scattering/haze slope rather than absorption by TiO/VO, with best-fit scattering enhancement and slope of 2.43$^\mathrm{+0.31}_\mathrm{-0.29}$ and --7.51$^\mathrm{+1.72}_\mathrm{-1.86}$ respectively. This is perhaps not all that surprising given the overall smoother, sloping shape of our transmission spectrum. For our best-fit model we reach a reduced $\chi^2$ of 1.485 (for 14 free parameters) which is comparable to the value in \citet{2018AJ....156..283E}. Interestingly, our retrieval also tries to fit the metal hydride CaH to the apparent feature at $\sim$0.7 microns, although we find only weak evidence for it ($\Delta$BIC = 2.14). CaH absorption is observed in some cool stars and brown dwarfs \citep[e.g.][]{1999ApJ...519..802K,2007A&A...473..245R,2011ASPC..448..531W}.
 
  The best-fit from our GMOS/WFC3 retrieval along with 1 sigma confidence contours are shown in Figure 6 and the full marginalised posterior probability densities are presented in Figure A3 in the appendix. A similar retrieval using PETRA was also carried out on the original STIS and WFC3 data and is presented in detail in \citet{2020arXiv201110626L}. In that study the Fe/H, H$_2$O and VO abundances were found to be slightly reduced from that reported in \citet{2018AJ....156..283E} though the overall interpretation remained largely the same. We also show the best-fit  model from this retrieval in Figure 6 for comparison.

 \begin{table}
    \caption{Retrieved atmospheric parameters using PETRA, where $\mathcal{U}(a,b)$ defines a unform distribution between $a$ and $b$.}
    \label{paramstable}
    \begin{tabular}{l|c|c}
        Parameter & Prior & Value \\
        \hline
        $\mathrm{log}(X_{\mathrm{H_2O}})$ & $\mathcal{U}(-12, -1)$ & $-5.05 ^{+ 0.19 }_{- 0.18 }$\\
        $\mathrm{log}(X_{\mathrm{TiO}})$ & $\mathcal{U}(-12, -1)$ & $-9.98 ^{+ 0.38 }_{- 0.65 }$\\
        $\mathrm{log}(X_{\mathrm{VO}})$ & $\mathcal{U}(-12, -1)$ & $-10.74 ^{+ 0.49 }_{- 0.56 }$\\
       
        $T_{0}$ & $\mathcal{U}(600, 4200)$  & $2664 ^{+ 120 }_{- 100 }$\\
        Scattering Enhancement Factor & $\mathcal{U}(-4, 8)$ & $2.43 ^{+ 0.31 }_{- 0.29 }$\\
        Scattering Slope & $\mathcal{U}(-15, 2)$  & $-7.51 ^{+ 1.72 }_{- 1.86 }$\\
        $\mathrm{log}(P_\mathrm{cloud})$[bar] & $\mathcal{U}(-4, 8)$ & $6.09 ^{+ 0.87 }_{- 0.79 }$\\
        \hline
    \end{tabular}
\end{table}

\section{Discussion}

WASP-121b is a close-in, ultra-hot Jupiter with a mass and radius of 1.18 \textit{M}$_\mathrm{J}$ and 1.7 \textit{R}$_\mathrm{J}$ respectively, an equilibrium temperature over 2400\,K and an orbital period of 1.27 days \citep{2016MNRAS.458.4025D}. In \citet{2017Natur.548...58E} strong evidence was found for the presence of a vertical thermal inversion using secondary eclipse measurements and this was subsequently confirmed in \citep{2020A&A...637A..36B}. The cause of such thermal inversions in the atmospheres of hot-Jupiters has traditionally been suggested to be due to absorption by TiO and/or VO \citep{2008ApJ...678.1419F}, though a convincing detection of either molecule has yet to be made for WASP-121b. \citet{2018AJ....156..283E} presented a STIS optical transmission spectrum of WASP-121b which showed features consistent with absorption by VO. However, follow-up secondary eclipse observations and recent observations at high-resolution have been unable to confirm this detection \citep{2020MNRAS.496.1638M,2020A&A...636A.117M}. This has led to some speculation that other species, such as the plethora of atomic metals including FeI that have recently been detected, might be the drivers for WASP-121b's temperature inversion \citep{2018ApJ...866...27L,2019MNRAS.485.5817G,2020MNRAS.493.2215G}. 

The transmission spectrum obtained from our joint analysis of two transits of WASP-121b is shown in Figure 6, with the STIS data of \citet{2018AJ....156..283E} and the results of a re-analysis of the ground-based photometry from \citet{2016MNRAS.458.4025D} (as presented in \citet{2016ApJ...822L...4E}) overplotted. For a small number of our points the values and uncertainties appear to deviate slightly from what would be expected from a simple weighted average of the two individual transits. This is likely due to more accurate limb darkening constraints from the joint fits, which should also improve the constraints on the corresponding systematics models. Overall, our transmission spectrum shows some agreement with that of \citet{2018AJ....156..283E} for wavelengths longward of $\sim$ 650\,nm, with some similar features being apparent in the middle portion of the spectrum. However, the results begin to diverge noticeably at wavelengths shorter than about $\sim$ 650\,nm, with the GMOS values being consistently larger than the STIS measurements across this short wavelength portion of the spectrum. We considered a number of factors that might be responsible for this apparent disagreement. The first of these is the treatment of limb darkening, which can potentially introduce small discrepancies in the transmission spectrum if not treated correctly, and will be particularly true at shorter wavelengths where the effects are strongest. For the analysis we have presented here we used the best-fit values calculated using PyLDTk and imposed wide Gaussian priors with a width of 0.1 to add some flexibility to the model fits. However, we also tried both fixing the limb darkening coefficients to the best-fit values and leaving them completely free in the fits, and found that this made almost no difference to our final transmission spectrum. In the \citet{2018AJ....156..283E} study they used 3D limb darkening coefficients from the STAGGER grid \citep{2013A&A...557A..26M}, so, as a check to ensure that our results were not overly dependent on our particular choice of limb darkening treatment, we re-extracted all light curves using identical bins as used for the G750L dataset in \citet{2018AJ....156..283E}, and repeated the analyses having fixed the limb darkening coefficients to the same best-fit values obtained from the STAGGER 3D stellar model. A comparison of our original spectrum and that produced from this subsequent analysis is shown in Figure A2 in the appendix. We find that adopting the STAGGER 3D limb darkening coefficients has a minimal influence on the overall shape of our transmission spectrum and is unable to explain the discrepancy between the GMOS and STIS datasets. This, together with the fact that we see little variation from fixing the limb darkening parameters or leaving them free, leads us to conclude that our results appear to be relatively insensitive to the treatment of limb darkening.

Stellar activity, including the possibility of unocculted spots or plages, has the potential to introduce both systematic offsets and wavelength-dependent biases in the measured transmission spectrum, with the effect being strongest for shorter wavelengths. In addition, changes in the wavelength dependence, as a result of time-variable spot coverages, could plausibly explain the observed discrepancy between the GMOS and STIS datasets given the separation in time of the individual observations. In \citet{2019AJ....157...96R} the estimated contamination for an F6 dwarf was found to be around a factor of $\sim$\,1.001, which corresponds to an offset in transit depth of $\sim$\,0.0025\,\%. The authors conclude that, whilst the effects of stellar contamination are more pronounced at shorter wavelengths, unocculted spots on typically active FGK dwarfs should only result in minor transit depth changes. In the WASP-121b discovery paper, \citet{2016MNRAS.458.4025D} used the 60 cm TRAPPIST telescope to investigate the photometric variability of WASP-121 over a six-week period, finding standard deviations of 1.6 mmag in the B band, 1.3 mmag in the V band and 1.1 mmag in the z$'$ band for the nightly photometry, and found no evidence for periodic variability above the $\sim$ 1 mmag level. Despite finding the photometry to be quiet, they nonetheless found that WASP-121 shows high scatter in its RV residuals, CCF bisector spans and FWHM from an analysis of CORALIE spectra, and suggested that this could point to the photosphere being plage dominated, since the lower flux ratios with respect to spots would result in smaller brightness variations \citep[e.g.][]{2014ApJ...796..132D}. 

Similarly, \citet{2018AJ....156..283E} also carried out photometric monitoring of WASP-121 using the Automated Imaging Telescope (AIT) during two separate campaigns in 2017 and 2018 and found a standard deviation about the yearly mean of 4.6 mmag for the 2017 campaign and 3.0 mmag for the 2018 campaign, with no significant periodicity detected for either campaign. They concluded that host star activity was unlikely to have significantly affected their transmission spectrum. In addition, they demonstrated that unocculted spots were unable to explain the shape of their measured spectrum under reasonable assumptions. Given this relatively quiet photometry and considering that both of our visits (separated by five days) analysed individually result in transmission spectra with very similar shapes, we likewise do not expect stellar activity to have significantly affected our results. However, taking into account the signatures of activity presented in the discovery paper, together with the temporal separation of the observations, we cannot entirely rule out the possibility that variable activity levels could at least partially be responsible for the discrepancies between our results and those obtained using STIS.

In order to quantify the potential effect of varying activity, we follow a similar approach to that outlined in \citet{2018AJ....156..283E}, which is an equivalent method to calculating spot correction factors as used in previous analyses \citep[e.g.][]{2011MNRAS.416.1443S, 2013MNRAS.434.3252H}. Using this approach we fit for the chromatic bias $\kappa(\lambda)$, which we here define as the difference between the measured GMOS transmission spectrum and the weighted average of the G750L STIS points and is given by:

\begin{equation}
   { \kappa(\lambda) = D \left(\frac{\alpha[1 - \beta(\lambda)]}{1 - \alpha[1 - \beta(\lambda)]}\right),}
\end{equation}

where {$\alpha$} is the fractional spot coverage (also known as the filling factor), $\beta$ is the wavelength-dependent spot-to-photosphere flux ratio and \textit{D} is the true transit depth in the absence of spots. In our fits we set \textit{D} equal to the weighted average of the STIS measurements and allow alpha to vary as a free parameter. This simple model assumes a fractional spot coverage and fixed spot temperature, and we use a model atmosphere obtained from the Phoenix grid \citep{2013A&A...553A...6H} with parameters \textit{T}$_\mathrm{\star}$ = 6500\,K, logg = 4.0 cgs, [Fe/H]= 0 dex to estimate the flux for the stellar photosphere. We use the same model to estimate the spot fluxes for a range of spot temperatures from 3500\,K up to 6000\,K and repeat our fits for each spot temperature. We found that none of our bias models were able to adequately explain the difference between the GMOS and STIS datasets, though the fits improved as the spot temperature was increased, however, this also required the spot fractions to increase from at least 2\% coverage to about 8\% coverage. We repeated this process subtracting the datasets in the opposite order, finding similar results. Given the suggestion in \citet{2016MNRAS.458.4025D} that the activity might be dominated by plage, we also tried fitting for higher temperature spots but found that a similar level of coverage was required in this case. This level of spot coverage variation would almost certainly be expected to result in larger features in the long-term monitoring provided by TRAPPIST, AIT and TESS which, although limited by both photometric precision and temporal coverage, suggest modulations no larger than about 5 mmag. Given the complexity involved in modelling the effects of stellar activity, we cannot completely rule out the possibility of changing activity levels with the simple model used here, although we consider it unlikely that it could account for the full variability observed between the data sets.

Another likely explanation for the disagreement is the possibility of remaining, unaccounted-for systematics in one or both of the GMOS transits, or within the STIS datasets, and perhaps even within both. Although independent analyses of our two individual transits already reveals an overall offset between them, we nonetheless recover transmission spectra with a consistent shape for both (see Figure 5), and we have shown that this does not change with our choice of priors for the transit shape parameters. It therefore seems strange that unaccounted-for systematics could conspire to alter the shape of our final transmission spectrum, unless they affected both individual transits in a similar way. However, this is equally true of the STIS datasets (which were analysed using a similar Gaussian process technique), where good agreement was found for separate fits to the individual G430L visits. We therefore cannot rule out the possibility that some unaccounted-for systematics in the GMOS and/or STIS transits are the cause of the discrepancy at short wavelengths. We further discuss potential causes of the disagreement between datasets in Sect.~\ref{sect:variability}.

Whilst we fail to detect any significant evidence for TiO or VO absorption in our PETRA retrieval, our transmission spectrum clearly shows excess absorption in the optical relative to that in the near-IR as reported in previous studies. Our results would therefore tend to support the hypothesis that other species are responsible for this optical excess. In particular, a number of atomic metals, including FeI, have already been detected in the atmospheres of WASP-121b and several other ultra-hot Jupiters at high-resolution \citep[e.g.][]{2019A&A...627A.165H,2019AJ....157...69C,2019AJ....158...91S,2020MNRAS.493.2215G,2020A&A...641A.123H,2020MNRAS.496..504N} and would be expected to be strong sources of optical absorption \citep{2008ApJ...678.1419F,2020ApJ...898L..14L}. The best-fit from our retrieval analysis favours a scattering/haze slope, and absorption by FeI and other atomic metals could potentially help to strengthen this slope. Additionally, UV observations of WASP-121b have shown evidence of atmospheric escape \citep{2019AJ....158...91S} and this could also enhance the effect of metals on the transmission spectrum if there is substantial material escaping from the atmosphere.

It has not been possible to uniquely identify these absorbing species in our low-resolution observations, though it seems likely that one or more of the reported species could be contributing to the observed optical excess and driving the thermal inversion \citep{2017Natur.548...58E,2020MNRAS.493.2215G,2020ApJ...898L..14L}. It is also possible that our non-detection of either TiO or VO could be explained simply by these molecules being condensed out on the cooler nightside and terminator regions. In Figure 7 we show a comparison of the GMOS transmission spectrum and WFC3 data from \citet{2016ApJ...822L...4E} with the predictions of a three-dimensional general circulation model (GCM) including TiO/VO/FeH opacity from \citet{2018A&A...617A.110P}. The two models correspond to a temperature map produced by a solar composition SPARC/MITgcm simulation with asymmetric temperatures between the west and east limbs and assume either a clear atmosphere or the presence of CaTiO$_3$ clouds. It is clear from this comparison that the GMOS data is in broad agreement with both the ground-based photometry and GCM predictions, even though the models have not themselves been fine-tuned to the data. Given this broad agreement we cannot entirely rule out TiO/VO despite not resolving such features in our transmission spectrum. Additionally, the reasonably strong haze slope favoured by PETRA is inconsistent with the short wavelength turn down in the GCM. There thus remains a number of plausible scenarios that could explain the current data and further observations at shorter wavelengths may help to determine if the absorption keeps rising (as favored by PETRA) or turns back down (as favored by the GCM).

A clearer picture of the overall atmospheric composition of WASP-121b may also emerge with broader wavelength coverage including in the IR with e.g. JWST. Accurate modelling of the full optical-infrared continuum is crucial for reliable retrievals, helping to break some of the degeneracies between model parameters \citep[]{2018MNRAS.480.5314P,2018AJ....155...29W}. A further way to break some of these degeneracies and help constrain the atmospheric properties could be achieved by combining the low-resolution observations with those obtained at high-resolution, either by joint fitting or by combining the posterior distributions of parameters, for example using direct likelihood evaluation techniques \citep[e.g.][]{2019AJ....157..114B,2020MNRAS.493.2215G}.

\subsection{Variability in the transmission spectrum?} 
\label{sect:variability}

\begin{figure}
\label{fig:parmentier}
 \includegraphics[width=1.07\columnwidth]{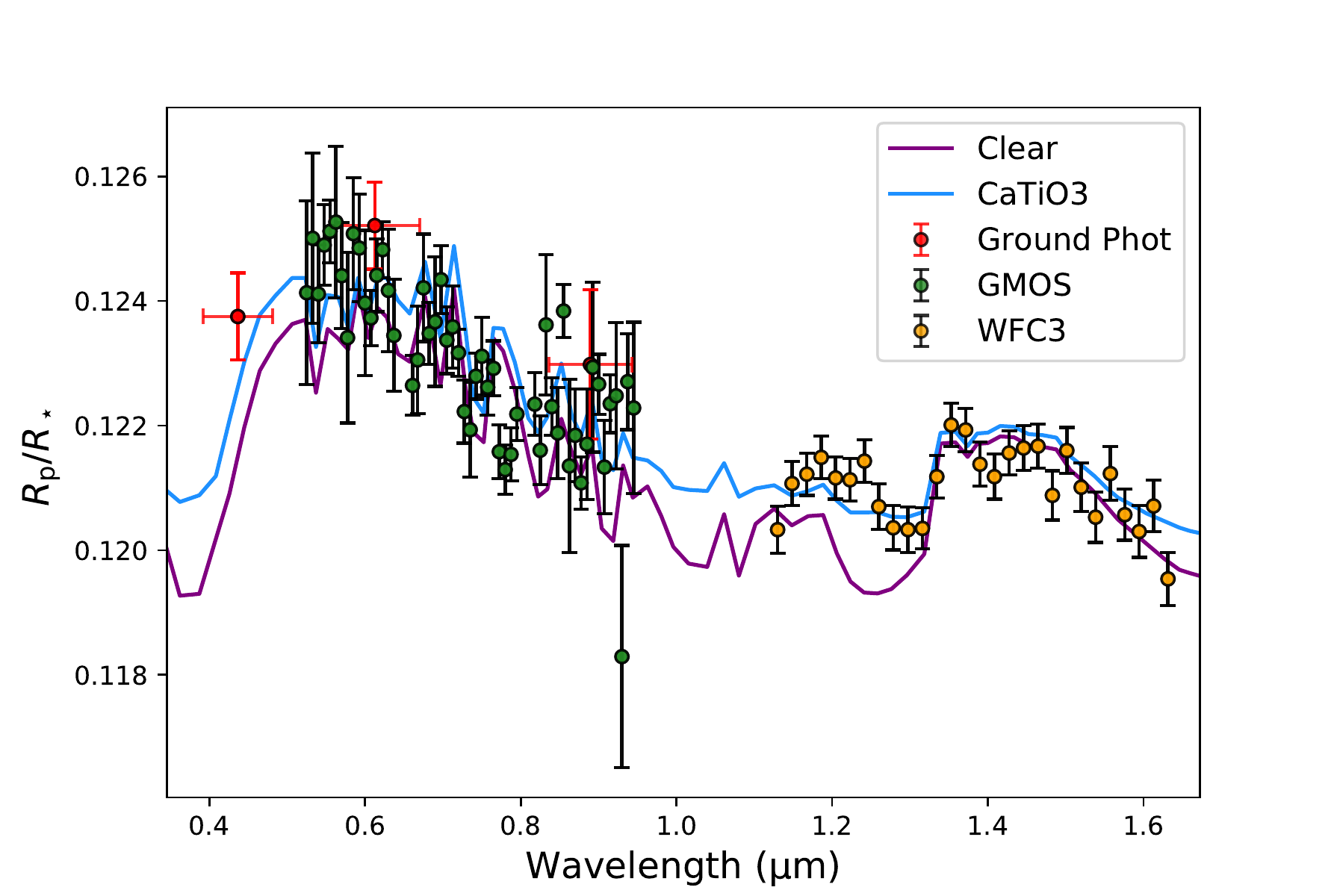}
 \caption{The GMOS/WFC3 data compared to the GCM predictions from \citet{2018A&A...617A.110P} with (blue) and without (purple) CaTiO$_3$ clouds. The ground-based photometry from \citet{2016MNRAS.458.4025D} is also shown.}
\end{figure}

A further, and somewhat intriguing, interpretation of the differences in the STIS and GMOS transmission spectra is that they are direct evidence of temporal variability in the atmospheric properties of WASP-121b, on timescales similar to that of the separation between the individual datasets. Such time-variability has previously been evoked to explain the changing phase curve offsets observed for HAT-P-7b, Kepler-76b and WASP-12b \citep{2016NatAs...1E...4A,2019AJ....157..239J,2019MNRAS.489.1995B}, and also the changing secondary eclipse depths for WASP-12b \citep{2019A&A...628A.115V,2019MNRAS.486.2397H}, although no such variations have yet been confirmed in transit observations.

In \citet{2013A&A...558A..91P}, the authors use a 3D GCM to simulate dynamical mixing and vertical settling for the well-studied hot-Jupiter HD 209458b, and show that strong variability of condensable chemical constituents is expected both spatially and temporally, driven mainly by the large day-to-night temperature contrast, and could result in observable variations during transit measurements. Furthermore, these results may also apply to silicate cloud/haze coverage, which could also display variability from epoch to epoch. Depending on the size of the condensates involved, the simulations predict an expected period of fifty to one hundred days for the largest amplitude variations, which is similar to the separation between the STIS/GMOS observations.

Similarly, in \citet{2020ApJ...888....2K}, the authors used GCM simulations to investigate the time-variability of hot-Jupiter atmospheres, finding $\sim$0.1\%–1\% variations in global-average, dayside-average, and nightside-average temperatures, and $\sim$1\%–10\% variations in globally averaged wind speeds, even when ignoring variability from magnetic effects and clouds. They also show how these variations could result in time-variable secondary eclipse depths, phase-curve amplitudes and offsets and terminator-averaged wind speeds. From phase-curve retrievals, WASP-121b is expected to have wind speeds of $\sim$7 km/s and a p-T profile which lies near the condensation curves of a number of species \citep[e.g.][]{2018A&A...617A.110P,2019arXiv190903000D}. It is therefore perhaps not all that surprising that small temperature fluctuations could result in significant spatial and temporal variations in atmospheric constituents and could lead to measurable variations in transit measurements.

Taken together the STIS/GMOS observations represent a somewhat unique dataset in that they both consist of sets of repeat observations separated by several days, with a longer period of $\sim$60 days between these sets. The STIS dataset comprises two repeat observations using the G430L grating and one with the G750L grating with individual analyses of the G430L data giving consistent results and showing a good level of agreement both with the NUV data from \citet{2019AJ....158...91S} and with the G750L dataset across the overlapping wavelength range. In their analysis \citet{2018AJ....156..283E} carried out a comprehensive range of tests to evaluate the robustness of their transmission spectrum, including a sensitivity analysis of their limb darkening treatment, the inclusion of additional GP inputs and a consideration of the effects of stellar activity. \citet{2018AJ....156..283E} also noticed a small discrepancy between their results and the ground-based photometry results obtained by \citet{2016MNRAS.458.4025D}, and also briefly speculated that intrinsic variability of the atmosphere from epoch to epoch could be responsible (these observations were separated by over 100 days). Both sets of individual observations for each instrument show excellent repeatability on short timescales, whilst the results over a longer period do not. It therefore seems plausible that the discrepancies between the STIS, GMOS and ground-based photometry could be the signature of time-variable weather conditions in the atmosphere of WASP-121b. However, we cannot entirely rule out the possibility that coherent systematics could be affecting the individual observations for one or both of the datasets in a similar way. Whilst a number of studies have shown some disagreement between ground/space-based instrumentation likely due to systematics \citep[e.g.][]{2017MNRAS.467.4591G,2019MNRAS.482..606G}, others have shown good agreement \citep[e.g.][]{2020AJ....160...51A} and we therefore consider temporal variability to represent a credible and intriguing possibility.

The fact that both the GMOS and STIS datasets can be fit with physically plausible PETRA models, are repeatable (over $\sim$\,day timescales), appear insensitive to limb darkening treatment and are a reasonable match to theoretical GCM predictions, means that we have no strong reason to discount the validity of either dataset and is therefore suggestive of the presence of time-variability.
However, neither can we fully rule out the possibility of more mundane causes, namely the influence of stellar activity and instrumental systematics, given the difficulty in robustly accounting for them even with our best instrumentation \citep[e.g.][]{2017MNRAS.467.4591G,2019MNRAS.482..606G,2019MNRAS.482.2065E}. 
We would therefore highly recommend further transit observations in the optical regime using appropriate timescales to specifically search for signs of such variable weather conditions.

\section{Conclusion}

We have presented ground-based Gemini/GMOS observations of the ultra-hot Jupiter WASP-121b covering two full transits and extracting a transmission spectrum over the wavelength range $\approx$\,500\,--\,950\,nm using the technique of differential spectrophotometry. We used a Gaussian process to simultaneously model the deterministic transit component and the instrumental systematics thereby avoiding the need to specify a specific functional form. We also introduced a new analysis technique for dealing with instrumental systematics in exoplanet transit light curves using Student’s-T processes, which can be thought of as a generalisation of the more commonly employed Gaussian processes. We used this new method to verify the analysis we have described here, finding that the results are in excellent agreement. We used the derived systematics models to correct our spectroscopic light curves by the application of common-mode corrections to improve the precision of our transmission spectrum.
In contrast to the STIS results, we find evidence for an increasing blueward slope and little evidence for absorption from either TiO or VO from an atmospheric retrieval using PETRA, in agreement with a number of recent studies performed at high-resolution. We suggest that this might point to other absorbers, such as some combination of atomic metals, being responsible for the optical excess and the vertical thermal inversion observed for WASP-121b. The scattering/haze slope favoured by our retrieval could be enhanced by both absorption from FeI and other atomic metals and from the atmospheric escape of material.

We considered a number of factors that might be responsible for the apparent discrepancy between the GMOS and STIS observations and conclude that, while it is still possible that the disagreement could be caused by either contamination from host star activity or some unaccounted-for systematic effects, the excellent repeatability of both individual datasets over day-long periods, and difference over longer month-long periods suggests that it is plausible that this discrepancy represents time-variable weather conditions in the atmosphere of WASP-121b, as predicted by theoretical models of ultra-hot Jupiters. We recommend further observations to both confirm this time-variability and help to further constrain the atmospheric properties, ultimately revealing the driver of the vertical thermal inversion.

\section*{Acknowledgements}
This work is Based on observations obtained at the international Gemini Observatory, a program of NSF’s NOIRLab, which is managed by the Association of Universities for Research in Astronomy (AURA) under a cooperative agreement with the National Science Foundation. on behalf of the Gemini Observatory partnership: the National Science Foundation (United States), National Research Council (Canada), Agencia Nacional de Investigaci\'{o}n y Desarrollo (Chile), Ministerio de Ciencia, Tecnolog\'{i}a e Innovaci\'{o}n (Argentina), Minist\'{e}rio da Ci\^{e}ncia, Tecnologia, Inova\c{c}\~{o}es e Comunica\c{c}\~{o}es (Brazil), and Korea Astronomy and Space Science Institute (Republic of Korea). This data was obtained under programme GS-2016B-Q-42. We thank Vivien Parmentier for providing the models shown in Fig.~\ref{fig:parmentier} from \citet{2018A&A...617A.110P}.
J.W. would like to acknowledge funding from the Northern Ireland Department for the Economy. N.P.G. gratefully acknowledges support from Science Foundation Ireland and the Royal Society in the form of a University Research Fellowship. We are grateful to the developers of the NumPy, SciPy, Matplotlib, iPython and Astropy packages, which were used extensively in this work \citep{2020SciPy-NMeth,2020NumPy-Array,4160251,4160265,2013A&A...558A..33A}.

\section*{Data availability}
The data in this article are available from the Gemini Observatory Archive (\url{https://archive.gemini.edu}) with program ID GS-2016B-Q-42. The data products generated from the raw data are available upon request from the author.

\bibliographystyle{mnras}
\bibliography{ref}

\begin{thebibliography}{}
\makeatletter
\relax
\def\mn@urlcharsother{\let\do\@makeother \do\$\do\&\do\#\do\^\do\_\do\%\do\~}
\def\mn@doi{\begingroup\mn@urlcharsother \@ifnextchar [ {\mn@doi@}
  {\mn@doi@[]}}
\def\mn@doi@[#1]#2{\def\@tempa{#1}\ifx\@tempa\@empty \href
  {http://dx.doi.org/#2} {doi:#2}\else \href {http://dx.doi.org/#2} {#1}\fi
  \endgroup}
\def\mn@eprint#1#2{\mn@eprint@#1:#2::\@nil}
\def\mn@eprint@arXiv#1{\href {http://arxiv.org/abs/#1} {{\tt arXiv:#1}}}
\def\mn@eprint@dblp#1{\href {http://dblp.uni-trier.de/rec/bibtex/#1.xml}
  {dblp:#1}}
\def\mn@eprint@#1:#2:#3:#4\@nil{\def\@tempa {#1}\def\@tempb {#2}\def\@tempc
  {#3}\ifx \@tempc \@empty \let \@tempc \@tempb \let \@tempb \@tempa \fi \ifx
  \@tempb \@empty \def\@tempb {arXiv}\fi \@ifundefined
  {mn@eprint@\@tempb}{\@tempb:\@tempc}{\expandafter \expandafter \csname
  mn@eprint@\@tempb\endcsname \expandafter{\@tempc}}}

\bibitem[\protect\citeauthoryear{{Alam} et~al.,}{{Alam}
  et~al.}{2020}]{2020AJ....160...51A}
{Alam} M.~K.,  et~al., 2020, \mn@doi [\aj] {10.3847/1538-3881/ab96cb}, \href
  {https://ui.adsabs.harvard.edu/abs/2020AJ....160...51A} {160, 51}

\bibitem[\protect\citeauthoryear{{Allard}, {Homeier}  \& {Freytag}}{{Allard}
  et~al.}{2011}]{2011ASPC..448...91A}
{Allard} F.,  {Homeier} D.,   {Freytag} B.,  2011, in {Johns-Krull} C.,
  {Browning} M.~K.,   {West} A.~A.,  eds,  Astronomical Society of the Pacific
  Conference Series Vol. 448, 16th Cambridge Workshop on Cool Stars, Stellar
  Systems, and the Sun. p.~91 (\mn@eprint {arXiv} {1011.5405})

\bibitem[\protect\citeauthoryear{{Armstrong}, {de Mooij}, {Barstow}, {Osborn},
  {Blake}  \& {Saniee}}{{Armstrong} et~al.}{2016}]{2016NatAs...1E...4A}
{Armstrong} D.~J.,  {de Mooij} E.,  {Barstow} J.,  {Osborn} H.~P.,  {Blake} J.,
    {Saniee} N.~F.,  2016, \mn@doi [Nature Astronomy]
  {10.1038/s41550-016-0004}, \href
  {https://ui.adsabs.harvard.edu/abs/2016NatAs...1E...4A} {1, 0004}

\bibitem[\protect\citeauthoryear{{Astropy Collaboration} et~al.,}{{Astropy
  Collaboration} et~al.}{2013}]{2013A&A...558A..33A}
{Astropy Collaboration} et~al., 2013, \mn@doi [\aap]
  {10.1051/0004-6361/201322068}, \href
  {https://ui.adsabs.harvard.edu/abs/2013A&A...558A..33A} {558, A33}

\bibitem[\protect\citeauthoryear{{Barman}, {Hauschildt}  \& {Allard}}{{Barman}
  et~al.}{2001}]{2001ApJ...556..885B}
{Barman} T.~S.,  {Hauschildt} P.~H.,   {Allard} F.,  2001, \mn@doi [\apj]
  {10.1086/321610}, \href
  {https://ui.adsabs.harvard.edu/abs/2001ApJ...556..885B} {556, 885}

\bibitem[\protect\citeauthoryear{{Barman}, {Macintosh}, {Konopacky}  \&
  {Marois}}{{Barman} et~al.}{2011}]{2011ApJ...733...65B}
{Barman} T.~S.,  {Macintosh} B.,  {Konopacky} Q.~M.,   {Marois} C.,  2011,
  \mn@doi [\apj] {10.1088/0004-637X/733/1/65}, \href
  {https://ui.adsabs.harvard.edu/abs/2011ApJ...733...65B} {733, 65}

\bibitem[\protect\citeauthoryear{{Bean} et~al.,}{{Bean}
  et~al.}{2011}]{2011ApJ...743...92B}
{Bean} J.~L.,  et~al., 2011, \mn@doi [\apj] {10.1088/0004-637X/743/1/92}, \href
  {https://ui.adsabs.harvard.edu/abs/2011ApJ...743...92B} {743, 92}

\bibitem[\protect\citeauthoryear{{Beatty}, {Madhusudhan}, {Tsiaras}, {Zhao},
  {Gilliland}, {Knutson}, {Shporer}  \& {Wright}}{{Beatty}
  et~al.}{2017}]{2017AJ....154..158B}
{Beatty} T.~G.,  {Madhusudhan} N.,  {Tsiaras} A.,  {Zhao} M.,  {Gilliland}
  R.~L.,  {Knutson} H.~A.,  {Shporer} A.,   {Wright} J.~T.,  2017, \mn@doi
  [\aj] {10.3847/1538-3881/aa899b}, \href
  {https://ui.adsabs.harvard.edu/abs/2017AJ....154..158B} {154, 158}

\bibitem[\protect\citeauthoryear{{Bell} et~al.,}{{Bell}
  et~al.}{2019}]{2019MNRAS.489.1995B}
{Bell} T.~J.,  et~al., 2019, \mn@doi [\mnras] {10.1093/mnras/stz2018}, \href
  {https://ui.adsabs.harvard.edu/abs/2019MNRAS.489.1995B} {489, 1995}

\bibitem[\protect\citeauthoryear{{Ben-Yami}, {Madhusudhan}, {Cabot},
  {Constantinou}, {Piette}, {Gandhi}  \& {Welbanks}}{{Ben-Yami}
  et~al.}{2020}]{2020ApJ...897L...5B}
{Ben-Yami} M.,  {Madhusudhan} N.,  {Cabot} S. H.~C.,  {Constantinou} S.,
  {Piette} A.,  {Gandhi} S.,   {Welbanks} L.,  2020, \mn@doi [\apjl]
  {10.3847/2041-8213/ab94aa}, \href
  {https://ui.adsabs.harvard.edu/abs/2020ApJ...897L...5B} {897, L5}

\bibitem[\protect\citeauthoryear{{Borsa} et~al.,}{{Borsa}
  et~al.}{2020}]{2020arXiv201101245B}
{Borsa} F.,  et~al., 2020, arXiv e-prints, \href
  {https://ui.adsabs.harvard.edu/abs/2020arXiv201101245B} {p. arXiv:2011.01245}

\bibitem[\protect\citeauthoryear{{Bourrier} et~al.,}{{Bourrier}
  et~al.}{2020a}]{2020A&A...635A.205B}
{Bourrier} V.,  et~al., 2020a, \mn@doi [\aap] {10.1051/0004-6361/201936640},
  \href {https://ui.adsabs.harvard.edu/abs/2020A&A...635A.205B} {635, A205}

\bibitem[\protect\citeauthoryear{{Bourrier} et~al.,}{{Bourrier}
  et~al.}{2020b}]{2020A&A...637A..36B}
{Bourrier} V.,  et~al., 2020b, \mn@doi [\aap] {10.1051/0004-6361/201936647},
  \href {https://ui.adsabs.harvard.edu/abs/2020A&A...637A..36B} {637, A36}

\bibitem[\protect\citeauthoryear{{Brogi} \& {Line}}{{Brogi} \&
  {Line}}{2019}]{2019AJ....157..114B}
{Brogi} M.,  {Line} M.~R.,  2019, \mn@doi [\aj] {10.3847/1538-3881/aaffd3},
  \href {https://ui.adsabs.harvard.edu/abs/2019AJ....157..114B} {157, 114}

\bibitem[\protect\citeauthoryear{{Brown}}{{Brown}}{2001}]{2001ApJ...553.1006B}
{Brown} T.~M.,  2001, \mn@doi [\apj] {10.1086/320950}, \href
  {https://ui.adsabs.harvard.edu/abs/2001ApJ...553.1006B} {553, 1006}

\bibitem[\protect\citeauthoryear{{Cabot}, {Madhusudhan}, {Welbanks}, {Piette}
  \& {Gandhi}}{{Cabot} et~al.}{2020}]{2020MNRAS.494..363C}
{Cabot} S. H.~C.,  {Madhusudhan} N.,  {Welbanks} L.,  {Piette} A.,   {Gandhi}
  S.,  2020, \mn@doi [\mnras] {10.1093/mnras/staa748}, \href
  {https://ui.adsabs.harvard.edu/abs/2020MNRAS.494..363C} {494, 363}

\bibitem[\protect\citeauthoryear{{Carter} et~al.,}{{Carter}
  et~al.}{2020}]{2020MNRAS.494.5449C}
{Carter} A.~L.,  et~al., 2020, \mn@doi [\mnras] {10.1093/mnras/staa1078}, \href
  {https://ui.adsabs.harvard.edu/abs/2020MNRAS.494.5449C} {494, 5449}

\bibitem[\protect\citeauthoryear{{Cauley}, {Shkolnik}, {Ilyin}, {Strassmeier},
  {Redfield}  \& {Jensen}}{{Cauley} et~al.}{2019}]{2019AJ....157...69C}
{Cauley} P.~W.,  {Shkolnik} E.~L.,  {Ilyin} I.,  {Strassmeier} K.~G.,
  {Redfield} S.,   {Jensen} A.,  2019, \mn@doi [\aj]
  {10.3847/1538-3881/aaf725}, \href
  {https://ui.adsabs.harvard.edu/abs/2019AJ....157...69C} {157, 69}

\bibitem[\protect\citeauthoryear{{Charbonneau}, {Brown}, {Noyes}  \&
  {Gilliland}}{{Charbonneau} et~al.}{2002}]{2002ApJ...568..377C}
{Charbonneau} D.,  {Brown} T.~M.,  {Noyes} R.~W.,   {Gilliland} R.~L.,  2002,
  \mn@doi [\apj] {10.1086/338770}, \href
  {https://ui.adsabs.harvard.edu/abs/2002ApJ...568..377C} {568, 377}

\bibitem[\protect\citeauthoryear{{Claret}}{{Claret}}{2000}]{2000A&A...363.1081C}
{Claret} A.,  2000, \aap, \href
  {https://ui.adsabs.harvard.edu/abs/2000A&A...363.1081C} {363, 1081}

\bibitem[\protect\citeauthoryear{{Daylan} et~al.,}{{Daylan}
  et~al.}{2019}]{2019arXiv190903000D}
{Daylan} T.,  et~al., 2019, arXiv e-prints, \href
  {https://ui.adsabs.harvard.edu/abs/2019arXiv190903000D} {p. arXiv:1909.03000}

\bibitem[\protect\citeauthoryear{{Delrez} et~al.,}{{Delrez}
  et~al.}{2016}]{2016MNRAS.458.4025D}
{Delrez} L.,  et~al., 2016, \mn@doi [\mnras] {10.1093/mnras/stw522}, \href
  {https://ui.adsabs.harvard.edu/abs/2016MNRAS.458.4025D} {458, 4025}

\bibitem[\protect\citeauthoryear{{Deming} et~al.,}{{Deming}
  et~al.}{2013}]{2013ApJ...774...95D}
{Deming} D.,  et~al., 2013, \mn@doi [\apj] {10.1088/0004-637X/774/2/95}, \href
  {https://ui.adsabs.harvard.edu/abs/2013ApJ...774...95D} {774, 95}

\bibitem[\protect\citeauthoryear{{Dumusque}, {Boisse}  \& {Santos}}{{Dumusque}
  et~al.}{2014}]{2014ApJ...796..132D}
{Dumusque} X.,  {Boisse} I.,   {Santos} N.~C.,  2014, \mn@doi [\apj]
  {10.1088/0004-637X/796/2/132}, \href
  {https://ui.adsabs.harvard.edu/abs/2014ApJ...796..132D} {796, 132}

\bibitem[\protect\citeauthoryear{{Espinoza} et~al.,}{{Espinoza}
  et~al.}{2019}]{2019MNRAS.482.2065E}
{Espinoza} N.,  et~al., 2019, \mn@doi [\mnras] {10.1093/mnras/sty2691}, \href
  {https://ui.adsabs.harvard.edu/abs/2019MNRAS.482.2065E} {482, 2065}

\bibitem[\protect\citeauthoryear{{Evans} et~al.,}{{Evans}
  et~al.}{2016}]{2016ApJ...822L...4E}
{Evans} T.~M.,  et~al., 2016, \mn@doi [\apjl] {10.3847/2041-8205/822/1/L4},
  \href {https://ui.adsabs.harvard.edu/abs/2016ApJ...822L...4E} {822, L4}

\bibitem[\protect\citeauthoryear{{Evans} et~al.,}{{Evans}
  et~al.}{2017}]{2017Natur.548...58E}
{Evans} T.~M.,  et~al., 2017, \mn@doi [\nat] {10.1038/nature23266}, \href
  {https://ui.adsabs.harvard.edu/abs/2017Natur.548...58E} {548, 58}

\bibitem[\protect\citeauthoryear{{Evans} et~al.,}{{Evans}
  et~al.}{2018}]{2018AJ....156..283E}
{Evans} T.~M.,  et~al., 2018, \mn@doi [\aj] {10.3847/1538-3881/aaebff}, \href
  {https://ui.adsabs.harvard.edu/abs/2018AJ....156..283E} {156, 283}

\bibitem[\protect\citeauthoryear{{Fortney}, {Lodders}, {Marley}  \&
  {Freedman}}{{Fortney} et~al.}{2008}]{2008ApJ...678.1419F}
{Fortney} J.~J.,  {Lodders} K.,  {Marley} M.~S.,   {Freedman} R.~S.,  2008,
  \mn@doi [\apj] {10.1086/528370}, \href
  {https://ui.adsabs.harvard.edu/abs/2008ApJ...678.1419F} {678, 1419}

\bibitem[\protect\citeauthoryear{{Gandhi} \& {Madhusudhan}}{{Gandhi} \&
  {Madhusudhan}}{2019}]{2019MNRAS.485.5817G}
{Gandhi} S.,  {Madhusudhan} N.,  2019, \mn@doi [\mnras] {10.1093/mnras/stz751},
  \href {https://ui.adsabs.harvard.edu/abs/2019MNRAS.485.5817G} {485, 5817}

\bibitem[\protect\citeauthoryear{Genz \& Bretz}{Genz \& Bretz}{2009}]{book}
Genz A.,  Bretz F.,  2009, Computation of Multivariate Normal and
  Probabilities.
~ Vol. 195, \mn@doi{10.1007/978-3-642-01689-9, }

\bibitem[\protect\citeauthoryear{{Gibson}, {Aigrain}, {Roberts}, {Evans},
  {Osborne}  \& {Pont}}{{Gibson} et~al.}{2012}]{2012MNRAS.419.2683G}
{Gibson} N.~P.,  {Aigrain} S.,  {Roberts} S.,  {Evans} T.~M.,  {Osborne} M.,
  {Pont} F.,  2012, \mn@doi [\mnras] {10.1111/j.1365-2966.2011.19915.x}, \href
  {https://ui.adsabs.harvard.edu/abs/2012MNRAS.419.2683G} {419, 2683}

\bibitem[\protect\citeauthoryear{{Gibson}, {Aigrain}, {Barstow}, {Evans},
  {Fletcher}  \& {Irwin}}{{Gibson} et~al.}{2013a}]{2013MNRAS.428.3680G}
{Gibson} N.~P.,  {Aigrain} S.,  {Barstow} J.~K.,  {Evans} T.~M.,  {Fletcher}
  L.~N.,   {Irwin} P.~G.~J.,  2013a, \mn@doi [\mnras] {10.1093/mnras/sts307},
  \href {https://ui.adsabs.harvard.edu/abs/2013MNRAS.428.3680G} {428, 3680}

\bibitem[\protect\citeauthoryear{{Gibson}, {Aigrain}, {Barstow}, {Evans},
  {Fletcher}  \& {Irwin}}{{Gibson} et~al.}{2013b}]{2013MNRAS.436.2974G}
{Gibson} N.~P.,  {Aigrain} S.,  {Barstow} J.~K.,  {Evans} T.~M.,  {Fletcher}
  L.~N.,   {Irwin} P.~G.~J.,  2013b, \mn@doi [\mnras] {10.1093/mnras/stt1783},
  \href {https://ui.adsabs.harvard.edu/abs/2013MNRAS.436.2974G} {436, 2974}

\bibitem[\protect\citeauthoryear{{Gibson}, {Nikolov}, {Sing}, {Barstow},
  {Evans}, {Kataria}  \& {Wilson}}{{Gibson} et~al.}{2017}]{2017MNRAS.467.4591G}
{Gibson} N.~P.,  {Nikolov} N.,  {Sing} D.~K.,  {Barstow} J.~K.,  {Evans} T.~M.,
   {Kataria} T.,   {Wilson} P.~A.,  2017, \mn@doi [\mnras]
  {10.1093/mnras/stx353}, \href
  {https://ui.adsabs.harvard.edu/abs/2017MNRAS.467.4591G} {467, 4591}

\bibitem[\protect\citeauthoryear{{Gibson}, {de Mooij}, {Evans}, {Merritt},
  {Nikolov}, {Sing}  \& {Watson}}{{Gibson} et~al.}{2019}]{2019MNRAS.482..606G}
{Gibson} N.~P.,  {de Mooij} E. J.~W.,  {Evans} T.~M.,  {Merritt} S.,  {Nikolov}
  N.,  {Sing} D.~K.,   {Watson} C.,  2019, \mn@doi [\mnras]
  {10.1093/mnras/sty2722}, \href
  {https://ui.adsabs.harvard.edu/abs/2019MNRAS.482..606G} {482, 606}

\bibitem[\protect\citeauthoryear{{Gibson} et~al.,}{{Gibson}
  et~al.}{2020}]{2020MNRAS.493.2215G}
{Gibson} N.~P.,  et~al., 2020, \mn@doi [\mnras] {10.1093/mnras/staa228}, \href
  {https://ui.adsabs.harvard.edu/abs/2020MNRAS.493.2215G} {493, 2215}

\bibitem[\protect\citeauthoryear{{Gillett}, {Low}  \& {Stein}}{{Gillett}
  et~al.}{1969}]{1969ApJ...157..925G}
{Gillett} F.~C.,  {Low} F.~J.,   {Stein} W.~A.,  1969, \mn@doi [\apj]
  {10.1086/150124}, \href
  {https://ui.adsabs.harvard.edu/abs/1969ApJ...157..925G} {157, 925}

\bibitem[\protect\citeauthoryear{{Gilliland} \& {Arribas}}{{Gilliland} \&
  {Arribas}}{2003}]{2003nicm.rept....1G}
{Gilliland} R.~L.,  {Arribas} S.,  2003, {High Signal-to-Noise Differential
  NICMOS Spectrophotometry}, Space Telescope NICMOS Instrument Science Report

\bibitem[\protect\citeauthoryear{{Gillon} et~al.,}{{Gillon}
  et~al.}{2012}]{2012A&A...542A...4G}
{Gillon} M.,  et~al., 2012, \mn@doi [\aap] {10.1051/0004-6361/201218817}, \href
  {https://ui.adsabs.harvard.edu/abs/2012A&A...542A...4G} {542, A4}

\bibitem[\protect\citeauthoryear{Harris et~al.,}{Harris
  et~al.}{2020}]{2020NumPy-Array}
Harris C.~R.,  et~al., 2020, \mn@doi [Nature] {10.1038/s41586-020-2649-2}, 585,
  357–362

\bibitem[\protect\citeauthoryear{{Hauschildt}, {Baron}  \&
  {Allard}}{{Hauschildt} et~al.}{1997}]{1997ApJ...483..390H}
{Hauschildt} P.~H.,  {Baron} E.,   {Allard} F.,  1997, \mn@doi [\apj]
  {10.1086/304233}, \href
  {https://ui.adsabs.harvard.edu/abs/1997ApJ...483..390H} {483, 390}

\bibitem[\protect\citeauthoryear{{Hauschildt}, {Allard}  \&
  {Baron}}{{Hauschildt} et~al.}{1999}]{1999ApJ...512..377H}
{Hauschildt} P.~H.,  {Allard} F.,   {Baron} E.,  1999, \mn@doi [\apj]
  {10.1086/306745}, \href
  {https://ui.adsabs.harvard.edu/abs/1999ApJ...512..377H} {512, 377}

\bibitem[\protect\citeauthoryear{{Herman}, {de Mooij}, {Jayawardhana}  \&
  {Brogi}}{{Herman} et~al.}{2020}]{2020AJ....160...93H}
{Herman} M.~K.,  {de Mooij} E. J.~W.,  {Jayawardhana} R.,   {Brogi} M.,  2020,
  \mn@doi [\aj] {10.3847/1538-3881/ab9e77}, \href
  {https://ui.adsabs.harvard.edu/abs/2020AJ....160...93H} {160, 93}

\bibitem[\protect\citeauthoryear{{Hoeijmakers} et~al.,}{{Hoeijmakers}
  et~al.}{2019}]{2019A&A...627A.165H}
{Hoeijmakers} H.~J.,  et~al., 2019, \mn@doi [\aap]
  {10.1051/0004-6361/201935089}, \href
  {https://ui.adsabs.harvard.edu/abs/2019A&A...627A.165H} {627, A165}

\bibitem[\protect\citeauthoryear{{Hoeijmakers} et~al.,}{{Hoeijmakers}
  et~al.}{2020}]{2020A&A...641A.123H}
{Hoeijmakers} H.~J.,  et~al., 2020, \mn@doi [\aap]
  {10.1051/0004-6361/202038365}, \href
  {https://ui.adsabs.harvard.edu/abs/2020A&A...641A.123H} {641, A123}

\bibitem[\protect\citeauthoryear{{Hook}, {J{\o}rgensen}, {Allington-Smith},
  {Davies}, {Metcalfe}, {Murowinski}  \& {Crampton}}{{Hook}
  et~al.}{2004}]{2004PASP..116..425H}
{Hook} I.~M.,  {J{\o}rgensen} I.,  {Allington-Smith} J.~R.,  {Davies} R.~L.,
  {Metcalfe} N.,  {Murowinski} R.~G.,   {Crampton} D.,  2004, \mn@doi [\pasp]
  {10.1086/383624}, \href
  {https://ui.adsabs.harvard.edu/abs/2004PASP..116..425H} {116, 425}

\bibitem[\protect\citeauthoryear{{Hooton}, {de Mooij}, {Watson}, {Gibson},
  {Galindo-Guil}, {Clavero}  \& {Merritt}}{{Hooton}
  et~al.}{2019}]{2019MNRAS.486.2397H}
{Hooton} M.~J.,  {de Mooij} E. J.~W.,  {Watson} C.~A.,  {Gibson} N.~P.,
  {Galindo-Guil} F.~J.,  {Clavero} R.,   {Merritt} S.~R.,  2019, \mn@doi
  [\mnras] {10.1093/mnras/stz966}, \href
  {https://ui.adsabs.harvard.edu/abs/2019MNRAS.486.2397H} {486, 2397}

\bibitem[\protect\citeauthoryear{{Hubeny}, {Burrows}  \& {Sudarsky}}{{Hubeny}
  et~al.}{2003}]{2003ApJ...594.1011H}
{Hubeny} I.,  {Burrows} A.,   {Sudarsky} D.,  2003, \mn@doi [\apj]
  {10.1086/377080}, \href
  {https://ui.adsabs.harvard.edu/abs/2003ApJ...594.1011H} {594, 1011}

\bibitem[\protect\citeauthoryear{{Huitson} et~al.,}{{Huitson}
  et~al.}{2013}]{2013MNRAS.434.3252H}
{Huitson} C.~M.,  et~al., 2013, \mn@doi [\mnras] {10.1093/mnras/stt1243}, \href
  {https://ui.adsabs.harvard.edu/abs/2013MNRAS.434.3252H} {434, 3252}

\bibitem[\protect\citeauthoryear{{Hunter}}{{Hunter}}{2007}]{4160265}
{Hunter} J.~D.,  2007, \mn@doi [Computing in Science Engineering]
  {10.1109/MCSE.2007.55}, 9, 90

\bibitem[\protect\citeauthoryear{{Husser}, {Wende-von Berg}, {Dreizler},
  {Homeier}, {Reiners}, {Barman}  \& {Hauschildt}}{{Husser}
  et~al.}{2013}]{2013A&A...553A...6H}
{Husser} T.~O.,  {Wende-von Berg} S.,  {Dreizler} S.,  {Homeier} D.,  {Reiners}
  A.,  {Barman} T.,   {Hauschildt} P.~H.,  2013, \mn@doi [\aap]
  {10.1051/0004-6361/201219058}, \href
  {https://ui.adsabs.harvard.edu/abs/2013A&A...553A...6H} {553, A6}

\bibitem[\protect\citeauthoryear{{Jackson}, {Adams}, {Sandidge}, {Kreyche}  \&
  {Briggs}}{{Jackson} et~al.}{2019}]{2019AJ....157..239J}
{Jackson} B.,  {Adams} E.,  {Sandidge} W.,  {Kreyche} S.,   {Briggs} J.,  2019,
  \mn@doi [\aj] {10.3847/1538-3881/ab1b30}, \href
  {https://ui.adsabs.harvard.edu/abs/2019AJ....157..239J} {157, 239}

\bibitem[\protect\citeauthoryear{{Kirkpatrick} et~al.,}{{Kirkpatrick}
  et~al.}{1999}]{1999ApJ...519..802K}
{Kirkpatrick} J.~D.,  et~al., 1999, \mn@doi [\apj] {10.1086/307414}, \href
  {https://ui.adsabs.harvard.edu/abs/1999ApJ...519..802K} {519, 802}

\bibitem[\protect\citeauthoryear{{Kitzmann} et~al.,}{{Kitzmann}
  et~al.}{2018}]{2018ApJ...863..183K}
{Kitzmann} D.,  et~al., 2018, \mn@doi [\apj] {10.3847/1538-4357/aace5a}, \href
  {https://ui.adsabs.harvard.edu/abs/2018ApJ...863..183K} {863, 183}

\bibitem[\protect\citeauthoryear{{Knutson}, {Howard}  \& {Isaacson}}{{Knutson}
  et~al.}{2010}]{2010ApJ...720.1569K}
{Knutson} H.~A.,  {Howard} A.~W.,   {Isaacson} H.,  2010, \mn@doi [\apj]
  {10.1088/0004-637X/720/2/1569}, \href
  {https://ui.adsabs.harvard.edu/abs/2010ApJ...720.1569K} {720, 1569}

\bibitem[\protect\citeauthoryear{{Komacek} \& {Showman}}{{Komacek} \&
  {Showman}}{2020}]{2020ApJ...888....2K}
{Komacek} T.~D.,  {Showman} A.~P.,  2020, \mn@doi [\apj]
  {10.3847/1538-4357/ab5b0b}, \href
  {https://ui.adsabs.harvard.edu/abs/2020ApJ...888....2K} {888, 2}

\bibitem[\protect\citeauthoryear{{Kreidberg} et~al.,}{{Kreidberg}
  et~al.}{2014}]{2014ApJ...793L..27K}
{Kreidberg} L.,  et~al., 2014, \mn@doi [\apjl] {10.1088/2041-8205/793/2/L27},
  \href {https://ui.adsabs.harvard.edu/abs/2014ApJ...793L..27K} {793, L27}

\bibitem[\protect\citeauthoryear{{Line} et~al.,}{{Line}
  et~al.}{2013}]{2013ApJ...775..137L}
{Line} M.~R.,  et~al., 2013, \mn@doi [\apj] {10.1088/0004-637X/775/2/137},
  \href {https://ui.adsabs.harvard.edu/abs/2013ApJ...775..137L} {775, 137}

\bibitem[\protect\citeauthoryear{{Lothringer} \& {Barman}}{{Lothringer} \&
  {Barman}}{2019}]{2019ApJ...876...69L}
{Lothringer} J.~D.,  {Barman} T.,  2019, \mn@doi [\apj]
  {10.3847/1538-4357/ab1485}, \href
  {https://ui.adsabs.harvard.edu/abs/2019ApJ...876...69L} {876, 69}

\bibitem[\protect\citeauthoryear{{Lothringer} \& {Barman}}{{Lothringer} \&
  {Barman}}{2020}]{2020AJ....159..289L}
{Lothringer} J.~D.,  {Barman} T.~S.,  2020, \mn@doi [\aj]
  {10.3847/1538-3881/ab8d33}, \href
  {https://ui.adsabs.harvard.edu/abs/2020AJ....159..289L} {159, 289}

\bibitem[\protect\citeauthoryear{{Lothringer}, {Barman}  \&
  {Koskinen}}{{Lothringer} et~al.}{2018}]{2018ApJ...866...27L}
{Lothringer} J.~D.,  {Barman} T.,   {Koskinen} T.,  2018, \mn@doi [\apj]
  {10.3847/1538-4357/aadd9e}, \href
  {https://ui.adsabs.harvard.edu/abs/2018ApJ...866...27L} {866, 27}

\bibitem[\protect\citeauthoryear{{Lothringer}, {Rustamkulov}, {Sing}, {Gibson},
  {Wilson}  \& {Schlaufman}}{{Lothringer} et~al.}{2020a}]{2020arXiv201110626L}
{Lothringer} J.~D.,  {Rustamkulov} Z.,  {Sing} D.~K.,  {Gibson} N.~P.,
  {Wilson} J.,   {Schlaufman} K.~C.,  2020a, arXiv e-prints, \href
  {https://ui.adsabs.harvard.edu/abs/2020arXiv201110626L} {p. arXiv:2011.10626}

\bibitem[\protect\citeauthoryear{{Lothringer}, {Fu}, {Sing}  \&
  {Barman}}{{Lothringer} et~al.}{2020b}]{2020ApJ...898L..14L}
{Lothringer} J.~D.,  {Fu} G.,  {Sing} D.~K.,   {Barman} T.~S.,  2020b, \mn@doi
  [\apjl] {10.3847/2041-8213/aba265}, \href
  {https://ui.adsabs.harvard.edu/abs/2020ApJ...898L..14L} {898, L14}

\bibitem[\protect\citeauthoryear{{MacDonald} \& {Madhusudhan}}{{MacDonald} \&
  {Madhusudhan}}{2017}]{2017MNRAS.469.1979M}
{MacDonald} R.~J.,  {Madhusudhan} N.,  2017, \mn@doi [\mnras]
  {10.1093/mnras/stx804}, \href
  {https://ui.adsabs.harvard.edu/abs/2017MNRAS.469.1979M} {469, 1979}

\bibitem[\protect\citeauthoryear{{Madhusudhan}, {Mousis}, {Johnson}  \&
  {Lunine}}{{Madhusudhan} et~al.}{2011}]{2011ApJ...743..191M}
{Madhusudhan} N.,  {Mousis} O.,  {Johnson} T.~V.,   {Lunine} J.~I.,  2011,
  \mn@doi [\apj] {10.1088/0004-637X/743/2/191}, \href
  {https://ui.adsabs.harvard.edu/abs/2011ApJ...743..191M} {743, 191}

\bibitem[\protect\citeauthoryear{{Magic}, {Collet}, {Asplund}, {Trampedach},
  {Hayek}, {Chiavassa}, {Stein}  \& {Nordlund}}{{Magic}
  et~al.}{2013}]{2013A&A...557A..26M}
{Magic} Z.,  {Collet} R.,  {Asplund} M.,  {Trampedach} R.,  {Hayek} W.,
  {Chiavassa} A.,  {Stein} R.~F.,   {Nordlund} {\r{A}}.,  2013, \mn@doi [\aap]
  {10.1051/0004-6361/201321274}, \href
  {https://ui.adsabs.harvard.edu/abs/2013A&A...557A..26M} {557, A26}

\bibitem[\protect\citeauthoryear{{Mandel} \& {Agol}}{{Mandel} \&
  {Agol}}{2002}]{2002ApJ...580L.171M}
{Mandel} K.,  {Agol} E.,  2002, \mn@doi [\apjl] {10.1086/345520}, \href
  {https://ui.adsabs.harvard.edu/abs/2002ApJ...580L.171M} {580, L171}

\bibitem[\protect\citeauthoryear{{Merritt} et~al.,}{{Merritt}
  et~al.}{2020}]{2020A&A...636A.117M}
{Merritt} S.~R.,  et~al., 2020, \mn@doi [\aap] {10.1051/0004-6361/201937409},
  \href {https://ui.adsabs.harvard.edu/abs/2020A&A...636A.117M} {636, A117}

\bibitem[\protect\citeauthoryear{{Mikal-Evans} et~al.,}{{Mikal-Evans}
  et~al.}{2019}]{2019MNRAS.488.2222M}
{Mikal-Evans} T.,  et~al., 2019, \mn@doi [\mnras] {10.1093/mnras/stz1753},
  \href {https://ui.adsabs.harvard.edu/abs/2019MNRAS.488.2222M} {488, 2222}

\bibitem[\protect\citeauthoryear{{Mikal-Evans}, {Sing}, {Kataria}, {Wakeford},
  {Mayne}, {Lewis}, {Barstow}  \& {Spake}}{{Mikal-Evans}
  et~al.}{2020}]{2020MNRAS.496.1638M}
{Mikal-Evans} T.,  {Sing} D.~K.,  {Kataria} T.,  {Wakeford} H.~R.,  {Mayne}
  N.~J.,  {Lewis} N.~K.,  {Barstow} J.~K.,   {Spake} J.~J.,  2020, \mn@doi
  [\mnras] {10.1093/mnras/staa1628}, \href
  {https://ui.adsabs.harvard.edu/abs/2020MNRAS.496.1638M} {496, 1638}

\bibitem[\protect\citeauthoryear{{Molli{\`e}re}, {van Boekel}, {Dullemond},
  {Henning}  \& {Mordasini}}{{Molli{\`e}re} et~al.}{2015}]{2015ApJ...813...47M}
{Molli{\`e}re} P.,  {van Boekel} R.,  {Dullemond} C.,  {Henning} T.,
  {Mordasini} C.,  2015, \mn@doi [\apj] {10.1088/0004-637X/813/1/47}, \href
  {https://ui.adsabs.harvard.edu/abs/2015ApJ...813...47M} {813, 47}

\bibitem[\protect\citeauthoryear{Nelder \& Mead}{Nelder \&
  Mead}{1965}]{10.1093/comjnl/7.4.308}
Nelder J.~A.,  Mead R.,  1965, \mn@doi [The Computer Journal]
  {10.1093/comjnl/7.4.308}, 7, 308

\bibitem[\protect\citeauthoryear{{Nikolov}, {Sing}, {Gibson}, {Fortney},
  {Evans}, {Barstow}, {Kataria}  \& {Wilson}}{{Nikolov}
  et~al.}{2016}]{2016ApJ...832..191N}
{Nikolov} N.,  {Sing} D.~K.,  {Gibson} N.~P.,  {Fortney} J.~J.,  {Evans} T.~M.,
   {Barstow} J.~K.,  {Kataria} T.,   {Wilson} P.~A.,  2016, \mn@doi [\apj]
  {10.3847/0004-637X/832/2/191}, \href
  {https://ui.adsabs.harvard.edu/abs/2016ApJ...832..191N} {832, 191}

\bibitem[\protect\citeauthoryear{{Nikolov} et~al.,}{{Nikolov}
  et~al.}{2018}]{2018Natur.557..526N}
{Nikolov} N.,  et~al., 2018, \mn@doi [\nat] {10.1038/s41586-018-0101-7}, \href
  {https://ui.adsabs.harvard.edu/abs/2018Natur.557..526N} {557, 526}

\bibitem[\protect\citeauthoryear{{Nugroho}, {Kawahara}, {Masuda}, {Hirano},
  {Kotani}  \& {Tajitsu}}{{Nugroho} et~al.}{2017}]{2017AJ....154..221N}
{Nugroho} S.~K.,  {Kawahara} H.,  {Masuda} K.,  {Hirano} T.,  {Kotani} T.,
  {Tajitsu} A.,  2017, \mn@doi [\aj] {10.3847/1538-3881/aa9433}, \href
  {https://ui.adsabs.harvard.edu/abs/2017AJ....154..221N} {154, 221}

\bibitem[\protect\citeauthoryear{{Nugroho}, {Gibson}, {de Mooij}, {Watson},
  {Kawahara}  \& {Merritt}}{{Nugroho} et~al.}{2020}]{2020MNRAS.496..504N}
{Nugroho} S.~K.,  {Gibson} N.~P.,  {de Mooij} E. J.~W.,  {Watson} C.~A.,
  {Kawahara} H.,   {Merritt} S.,  2020, \mn@doi [\mnras]
  {10.1093/mnras/staa1459}, \href
  {https://ui.adsabs.harvard.edu/abs/2020MNRAS.496..504N} {496, 504}

\bibitem[\protect\citeauthoryear{{Parmentier} \& {Guillot}}{{Parmentier} \&
  {Guillot}}{2014}]{2014A&A...562A.133P}
{Parmentier} V.,  {Guillot} T.,  2014, \mn@doi [\aap]
  {10.1051/0004-6361/201322342}, \href
  {https://ui.adsabs.harvard.edu/abs/2014A&A...562A.133P} {562, A133}

\bibitem[\protect\citeauthoryear{{Parmentier}, {Showman}  \&
  {Lian}}{{Parmentier} et~al.}{2013}]{2013A&A...558A..91P}
{Parmentier} V.,  {Showman} A.~P.,   {Lian} Y.,  2013, \mn@doi [\aap]
  {10.1051/0004-6361/201321132}, \href
  {https://ui.adsabs.harvard.edu/abs/2013A&A...558A..91P} {558, A91}

\bibitem[\protect\citeauthoryear{{Parmentier} et~al.,}{{Parmentier}
  et~al.}{2018}]{2018A&A...617A.110P}
{Parmentier} V.,  et~al., 2018, \mn@doi [\aap] {10.1051/0004-6361/201833059},
  \href {https://ui.adsabs.harvard.edu/abs/2018A&A...617A.110P} {617, A110}

\bibitem[\protect\citeauthoryear{{Parviainen} \& {Aigrain}}{{Parviainen} \&
  {Aigrain}}{2015}]{2015MNRAS.453.3821P}
{Parviainen} H.,  {Aigrain} S.,  2015, \mn@doi [\mnras]
  {10.1093/mnras/stv1857}, \href
  {https://ui.adsabs.harvard.edu/abs/2015MNRAS.453.3821P} {453, 3821}

\bibitem[\protect\citeauthoryear{{Perez} \& {Granger}}{{Perez} \&
  {Granger}}{2007}]{4160251}
{Perez} F.,  {Granger} B.~E.,  2007, \mn@doi [Computing in Science Engineering]
  {10.1109/MCSE.2007.53}, 9, 21

\bibitem[\protect\citeauthoryear{{Pinhas}, {Rackham}, {Madhusudhan}  \&
  {Apai}}{{Pinhas} et~al.}{2018}]{2018MNRAS.480.5314P}
{Pinhas} A.,  {Rackham} B.~V.,  {Madhusudhan} N.,   {Apai} D.,  2018, \mn@doi
  [\mnras] {10.1093/mnras/sty2209}, \href
  {https://ui.adsabs.harvard.edu/abs/2018MNRAS.480.5314P} {480, 5314}

\bibitem[\protect\citeauthoryear{{Pino} et~al.,}{{Pino}
  et~al.}{2020}]{2020ApJ...894L..27P}
{Pino} L.,  et~al., 2020, \mn@doi [\apjl] {10.3847/2041-8213/ab8c44}, \href
  {https://ui.adsabs.harvard.edu/abs/2020ApJ...894L..27P} {894, L27}

\bibitem[\protect\citeauthoryear{{Pont} et~al.,}{{Pont}
  et~al.}{2007}]{2007A&A...476.1347P}
{Pont} F.,  et~al., 2007, \mn@doi [\aap] {10.1051/0004-6361:20078269}, \href
  {https://ui.adsabs.harvard.edu/abs/2007A&A...476.1347P} {476, 1347}

\bibitem[\protect\citeauthoryear{{Pont}, {Sing}, {Gibson}, {Aigrain}, {Henry}
  \& {Husnoo}}{{Pont} et~al.}{2013}]{2013MNRAS.432.2917P}
{Pont} F.,  {Sing} D.~K.,  {Gibson} N.~P.,  {Aigrain} S.,  {Henry} G.,
  {Husnoo} N.,  2013, \mn@doi [\mnras] {10.1093/mnras/stt651}, \href
  {https://ui.adsabs.harvard.edu/abs/2013MNRAS.432.2917P} {432, 2917}

\bibitem[\protect\citeauthoryear{{Rackham}, {Apai}  \& {Giampapa}}{{Rackham}
  et~al.}{2019}]{2019AJ....157...96R}
{Rackham} B.~V.,  {Apai} D.,   {Giampapa} M.~S.,  2019, \mn@doi [\aj]
  {10.3847/1538-3881/aaf892}, \href
  {https://ui.adsabs.harvard.edu/abs/2019AJ....157...96R} {157, 96}

\bibitem[\protect\citeauthoryear{Rasmussen \& Williams}{Rasmussen \&
  Williams}{2005}]{10.5555/1162254}
Rasmussen C.~E.,  Williams C. K.~I.,  2005, Gaussian Processes for Machine
  Learning (Adaptive Computation and Machine Learning).
The MIT Press

\bibitem[\protect\citeauthoryear{{Reiners}, {Homeier}, {Hauschildt}  \&
  {Allard}}{{Reiners} et~al.}{2007}]{2007A&A...473..245R}
{Reiners} A.,  {Homeier} D.,  {Hauschildt} P.~H.,   {Allard} F.,  2007, \mn@doi
  [\aap] {10.1051/0004-6361:20077963}, \href
  {https://ui.adsabs.harvard.edu/abs/2007A&A...473..245R} {473, 245}

\bibitem[\protect\citeauthoryear{{Ridgway}}{{Ridgway}}{1974}]{1974ApJ...187L..41R}
{Ridgway} S.~T.,  1974, \mn@doi [\apjl] {10.1086/181388}, \href
  {https://ui.adsabs.harvard.edu/abs/1974ApJ...187L..41R} {187, L41}

\bibitem[\protect\citeauthoryear{{Salz}, {Schneider}, {Fossati}, {Czesla},
  {France}  \& {Schmitt}}{{Salz} et~al.}{2019}]{2019A&A...623A..57S}
{Salz} M.,  {Schneider} P.~C.,  {Fossati} L.,  {Czesla} S.,  {France} K.,
  {Schmitt} J.~H.~M.~M.,  2019, \mn@doi [\aap] {10.1051/0004-6361/201732419},
  \href {https://ui.adsabs.harvard.edu/abs/2019A&A...623A..57S} {623, A57}

\bibitem[\protect\citeauthoryear{{Seager} \& {Sasselov}}{{Seager} \&
  {Sasselov}}{2000}]{2000ApJ...537..916S}
{Seager} S.,  {Sasselov} D.~D.,  2000, \mn@doi [\apj] {10.1086/309088}, \href
  {https://ui.adsabs.harvard.edu/abs/2000ApJ...537..916S} {537, 916}

\bibitem[\protect\citeauthoryear{{Sedaghati} et~al.,}{{Sedaghati}
  et~al.}{2017}]{2017Natur.549..238S}
{Sedaghati} E.,  et~al., 2017, \mn@doi [\nat] {10.1038/nature23651}, \href
  {https://ui.adsabs.harvard.edu/abs/2017Natur.549..238S} {549, 238}

\bibitem[\protect\citeauthoryear{Shah}{Shah}{2013}]{Shah2013BayesianOU}
Shah A.,  2013.

\bibitem[\protect\citeauthoryear{Shah, Wilson  \& Ghahramani}{Shah
  et~al.}{2014}]{pmlr-v33-shah14}
Shah A.,  Wilson A.,   Ghahramani Z.,  2014. PMLR, Reykjavik, Iceland, pp
  877--885, \url {http://proceedings.mlr.press/v33/shah14.html}

\bibitem[\protect\citeauthoryear{{Sing} et~al.,}{{Sing}
  et~al.}{2011}]{2011MNRAS.416.1443S}
{Sing} D.~K.,  et~al., 2011, \mn@doi [\mnras]
  {10.1111/j.1365-2966.2011.19142.x}, \href
  {https://ui.adsabs.harvard.edu/abs/2011MNRAS.416.1443S} {416, 1443}

\bibitem[\protect\citeauthoryear{{Sing} et~al.,}{{Sing}
  et~al.}{2015}]{2015MNRAS.446.2428S}
{Sing} D.~K.,  et~al., 2015, \mn@doi [\mnras] {10.1093/mnras/stu2279}, \href
  {https://ui.adsabs.harvard.edu/abs/2015MNRAS.446.2428S} {446, 2428}

\bibitem[\protect\citeauthoryear{{Sing} et~al.,}{{Sing}
  et~al.}{2016}]{2016Natur.529...59S}
{Sing} D.~K.,  et~al., 2016, \mn@doi [\nat] {10.1038/nature16068}, \href
  {https://ui.adsabs.harvard.edu/abs/2016Natur.529...59S} {529, 59}

\bibitem[\protect\citeauthoryear{{Sing} et~al.,}{{Sing}
  et~al.}{2019}]{2019AJ....158...91S}
{Sing} D.~K.,  et~al., 2019, \mn@doi [\aj] {10.3847/1538-3881/ab2986}, \href
  {https://ui.adsabs.harvard.edu/abs/2019AJ....158...91S} {158, 91}

\bibitem[\protect\citeauthoryear{{Spake} et~al.,}{{Spake}
  et~al.}{2018}]{2018Natur.557...68S}
{Spake} J.~J.,  et~al., 2018, \mn@doi [\nat] {10.1038/s41586-018-0067-5}, \href
  {https://ui.adsabs.harvard.edu/abs/2018Natur.557...68S} {557, 68}

\bibitem[\protect\citeauthoryear{{Stevenson} et~al.,}{{Stevenson}
  et~al.}{2010}]{2010Natur.464.1161S}
{Stevenson} K.~B.,  et~al., 2010, \mn@doi [\nat] {10.1038/nature09013}, \href
  {https://ui.adsabs.harvard.edu/abs/2010Natur.464.1161S} {464, 1161}

\bibitem[\protect\citeauthoryear{{Stevenson}, {Bean}, {Seifahrt}, {D{\'e}sert},
  {Madhusudhan}, {Bergmann}, {Kreidberg}  \& {Homeier}}{{Stevenson}
  et~al.}{2014}]{2014AJ....147..161S}
{Stevenson} K.~B.,  {Bean} J.~L.,  {Seifahrt} A.,  {D{\'e}sert} J.-M.,
  {Madhusudhan} N.,  {Bergmann} M.,  {Kreidberg} L.,   {Homeier} D.,  2014,
  \mn@doi [\aj] {10.1088/0004-6256/147/6/161}, \href
  {https://ui.adsabs.harvard.edu/abs/2014AJ....147..161S} {147, 161}

\bibitem[\protect\citeauthoryear{{Swain}, {Vasisht}, {Tinetti}, {Bouwman},
  {Chen}, {Yung}, {Deming}  \& {Deroo}}{{Swain}
  et~al.}{2009}]{2009ApJ...690L.114S}
{Swain} M.~R.,  {Vasisht} G.,  {Tinetti} G.,  {Bouwman} J.,  {Chen} P.,  {Yung}
  Y.,  {Deming} D.,   {Deroo} P.,  2009, \mn@doi [\apjl]
  {10.1088/0004-637X/690/2/L114}, \href
  {https://ui.adsabs.harvard.edu/abs/2009ApJ...690L.114S} {690, L114}

\bibitem[\protect\citeauthoryear{Ter~Braak}{Ter~Braak}{2006}]{cite-key}
Ter~Braak C. J. F.~T.,  2006, \mn@doi [Statistics and Computing]
  {10.1007/s11222-006-8769-1}, 16, 239

\bibitem[\protect\citeauthoryear{Ter~Braak \& Vrugt}{Ter~Braak \&
  Vrugt}{2008}]{cite-key2}
Ter~Braak C. J.~F.,  Vrugt J.~A.,  2008, \mn@doi [Statistics and Computing]
  {10.1007/s11222-008-9104-9}, 18, 435

\bibitem[\protect\citeauthoryear{Tracey \& Wolpert}{Tracey \&
  Wolpert}{2018}]{doi:10.2514/6.2018-1659}
Tracey B.~D.,  Wolpert D.,  2018, Upgrading from Gaussian Processes to
  Student’s-T Processes.
 (\mn@eprint {} {https://arc.aiaa.org/doi/pdf/10.2514/6.2018-1659}),
  \mn@doi{10.2514/6.2018-1659}, \url
  {https://arc.aiaa.org/doi/abs/10.2514/6.2018-1659}

\bibitem[\protect\citeauthoryear{Virtanen et~al.,}{Virtanen
  et~al.}{2020}]{2020SciPy-NMeth}
Virtanen P.,  et~al., 2020, \mn@doi [Nature Methods]
  {10.1038/s41592-019-0686-2}, \href {https://rdcu.be/b08Wh} {17, 261}

\bibitem[\protect\citeauthoryear{{Wakeford} et~al.,}{{Wakeford}
  et~al.}{2018}]{2018AJ....155...29W}
{Wakeford} H.~R.,  et~al., 2018, \mn@doi [\aj] {10.3847/1538-3881/aa9e4e},
  \href {https://ui.adsabs.harvard.edu/abs/2018AJ....155...29W} {155, 29}

\bibitem[\protect\citeauthoryear{{Wallace}, {Prather}  \& {Belton}}{{Wallace}
  et~al.}{1974}]{1974ApJ...193..481W}
{Wallace} L.,  {Prather} M.,   {Belton} M.~J.~S.,  1974, \mn@doi [\apj]
  {10.1086/153184}, \href
  {https://ui.adsabs.harvard.edu/abs/1974ApJ...193..481W} {193, 481}

\bibitem[\protect\citeauthoryear{{West}, {Bochanski}, {Bowler}, {Dotter},
  {Johnson}, {L{\'e}pine}, {Rojas-Ayala}  \& {Schweitzer}}{{West}
  et~al.}{2011}]{2011ASPC..448..531W}
{West} A.~A.,  {Bochanski} J.~J.,  {Bowler} B.~P.,  {Dotter} A.,  {Johnson}
  J.~A.,  {L{\'e}pine} S.,  {Rojas-Ayala} B.,   {Schweitzer} A.,  2011, in
  {Johns-Krull} C.,  {Browning} M.~K.,   {West} A.~A.,  eds,  Astronomical
  Society of the Pacific Conference Series Vol. 448, 16th Cambridge Workshop on
  Cool Stars, Stellar Systems, and the Sun. p.~531 (\mn@eprint {arXiv}
  {1101.1086})

\bibitem[\protect\citeauthoryear{{Wilson} et~al.,}{{Wilson}
  et~al.}{2020}]{2020MNRAS.497.5155W}
{Wilson} J.,  et~al., 2020, \mn@doi [\mnras] {10.1093/mnras/staa2307}, \href
  {https://ui.adsabs.harvard.edu/abs/2020MNRAS.497.5155W} {497, 5155}

\bibitem[\protect\citeauthoryear{{von Essen} et~al.,}{{von Essen}
  et~al.}{2019}]{2019A&A...628A.115V}
{von Essen} C.,  et~al., 2019, \mn@doi [\aap] {10.1051/0004-6361/201935312},
  \href {https://ui.adsabs.harvard.edu/abs/2019A&A...628A.115V} {628, A115}

\makeatother
\end{thebibliography}

\appendix

\section{}

\begin{figure*}
 \includegraphics[width=\textwidth]{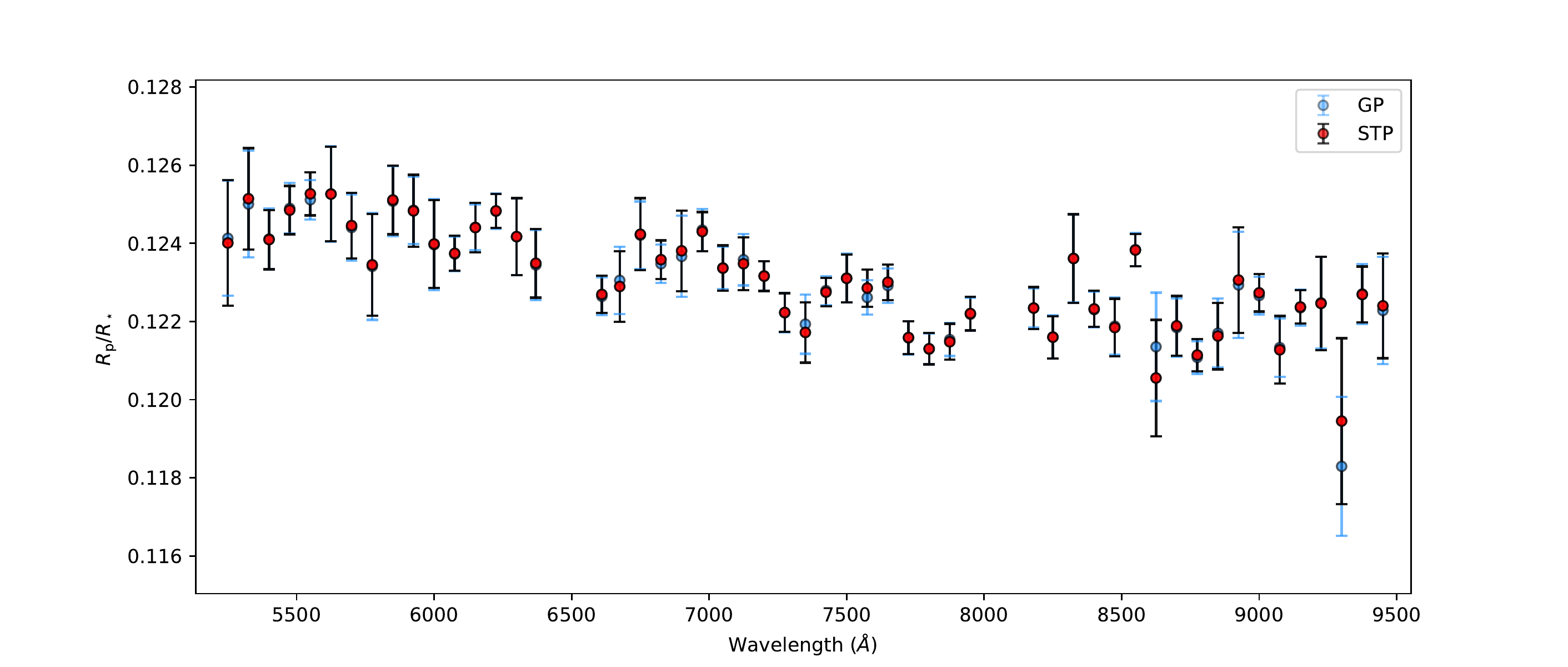}
 \caption{GMOS transmission spectrum of WASP-121b recovered from our GP and STP analyses. The blue points are the results of the GP analysis, whilst the red points are the result of the STP analysis. Both analyses used the same priors for the white light curves.}
 \label{fig:fors_spec}
\end{figure*}

\begin{figure*}
 \includegraphics[width=\textwidth]{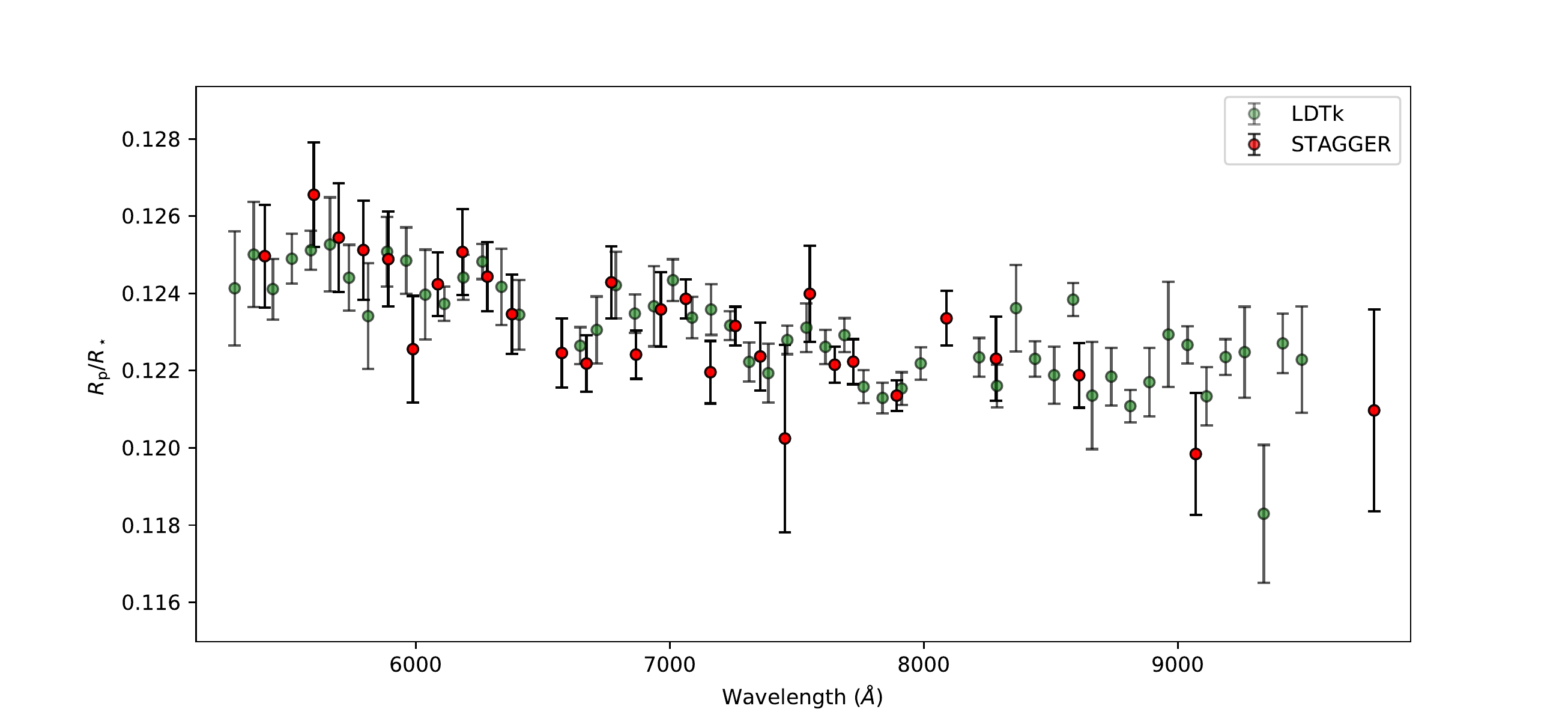}
 \caption{Our original GMOS transmission spectrum of WASP-121b compared with the result of a re-analysis using identical bins and the same STAGGER 3D limb darkening coefficients as in \citet{2018AJ....156..283E}. The green points show our original transmission spectrum whilst the red points are the result of the re-analysis.}
 \label{fig:fors_spec}
\end{figure*}

\begin{figure*}
 \includegraphics[width=\textwidth]{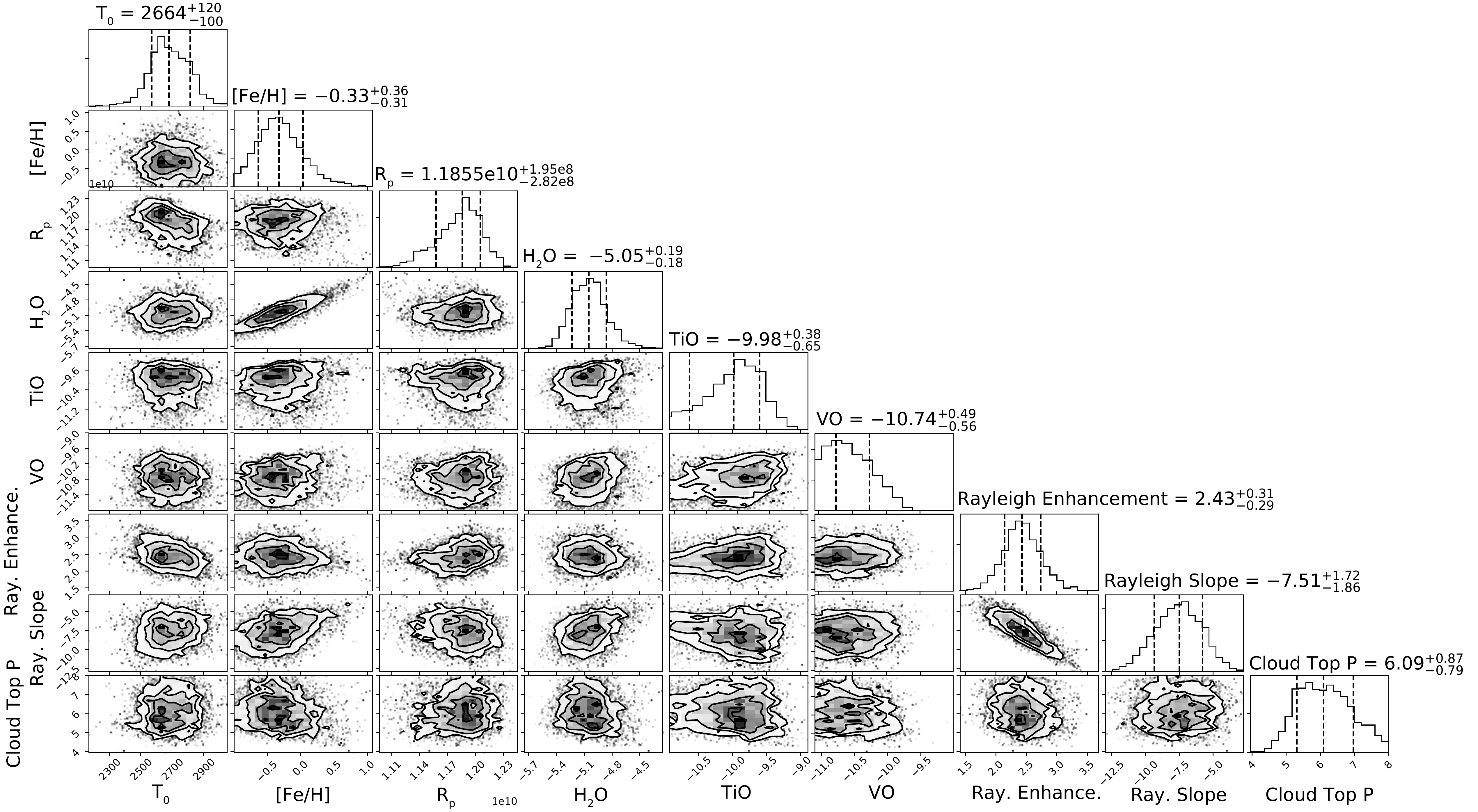}
 \caption{Marginalised posterior probability densities for the atmospheric retrieval performed using the PETRA code on the combined optical-infrared transmission spectrum of WASP-121b.}
\end{figure*}

\bsp	
\label{lastpage}
\end{document}